%% file: main.tex
\documentclass[sigconf]{acmart}

\usepackage{graphicx}
\usepackage{booktabs}
\usepackage{multirow}
\usepackage{colortbl}
\usepackage{xspace}
\usepackage{graphicx}
\usepackage{xcolor}
\usepackage{subcaption}
\usepackage{soul}
\usepackage{fontawesome}
\usepackage{algpseudocode}
\usepackage{algorithm}
\usepackage{dirtytalk}
\usepackage{listings}
\usepackage{float}
\usepackage{caption}
\usepackage{balance}

\input{00-macro}

\definecolor{commentcolor}{rgb}{1, 0.25, 0.55}

\newcommand{\toolname}{\textsc{Mocha}}

\setcopyright{cc}
\setcctype{by}
\copyrightyear{2025}
\acmYear{2025}
\acmDOI{10.1145/3706598.3713708}
\acmISBN{979-8-4007-1394-1/25/04}
\acmConference[CHI '25]{CHI Conference on Human Factors in Computing Systems}{April 26--May 01, 2025}{Yokohama, Japan}
\acmBooktitle{CHI Conference on Human Factors in Computing Systems (CHI '25), April 26--May 01, 2025, Yokohama, Japan}

\begin{document}
\title{Supporting Co-Adaptive Machine Teaching through Human Concept Learning and Cognitive Theories}

\author{Simret Araya Gebreegziabher}
\affiliation{%
\department{Department of Computer Science and Engineering} 
  \institution{University of Notre Dame}
  \city{Notre Dame}
  \state{IN}
  \country{USA}}
\email{sgebreeg@nd.edu}

\author{Yukun Yang}
\affiliation{%
\department{Department of Computer Science and Engineering} 
  \institution{University of Notre Dame}
  \city{Notre Dame}
  \state{IN}
  \country{USA}}
\email{yyang35@nd.edu}

\author{Elena L. Glassman}
\authornote{Co-senior authors contributed equally.}
\affiliation{%
\department{School of Engineering and Applied Sciences} 
  \institution{Harvard University}
  \city{Cambridge}
  \state{MA}
  \country{USA}}
\email{glassman@seas.harvard.edu}

\author{Toby Jia-Jun Li}
\authornotemark[1]
\affiliation{%
\department{Department of Computer Science and Engineering} 
  \institution{University of Notre Dame}
  \city{Notre Dame}
  \state{IN}
  \country{USA}}
\email{toby.j.li@nd.edu }

\begin{abstract}

An important challenge in interactive machine learning, particularly in subjective or ambiguous domains, is fostering bi-directional alignment between humans and models. Users teach models their concept definition through data labeling, while refining their own understandings throughout the process. To facilitate this, we introduce \toolname, an interactive machine learning tool informed by two theories of human concept learning and cognition. First, it utilizes a neuro-symbolic pipeline to support Variation Theory-based counterfactual data generation. By asking users to annotate counterexamples that are syntactically and semantically similar to already-annotated data but predicted to have different labels, the system can learn more effectively while helping users understand the model and reflect on their own label definitions. Second, \toolname{} uses Structural Alignment Theory to present groups of counterexamples, helping users comprehend alignable differences between data items and annotate them in batch. We validated \toolname’s effectiveness and usability through a lab study with 18 participants.

\end{abstract}




\begin{teaserfigure}
    \centering
    \includegraphics[width=0.8\linewidth]{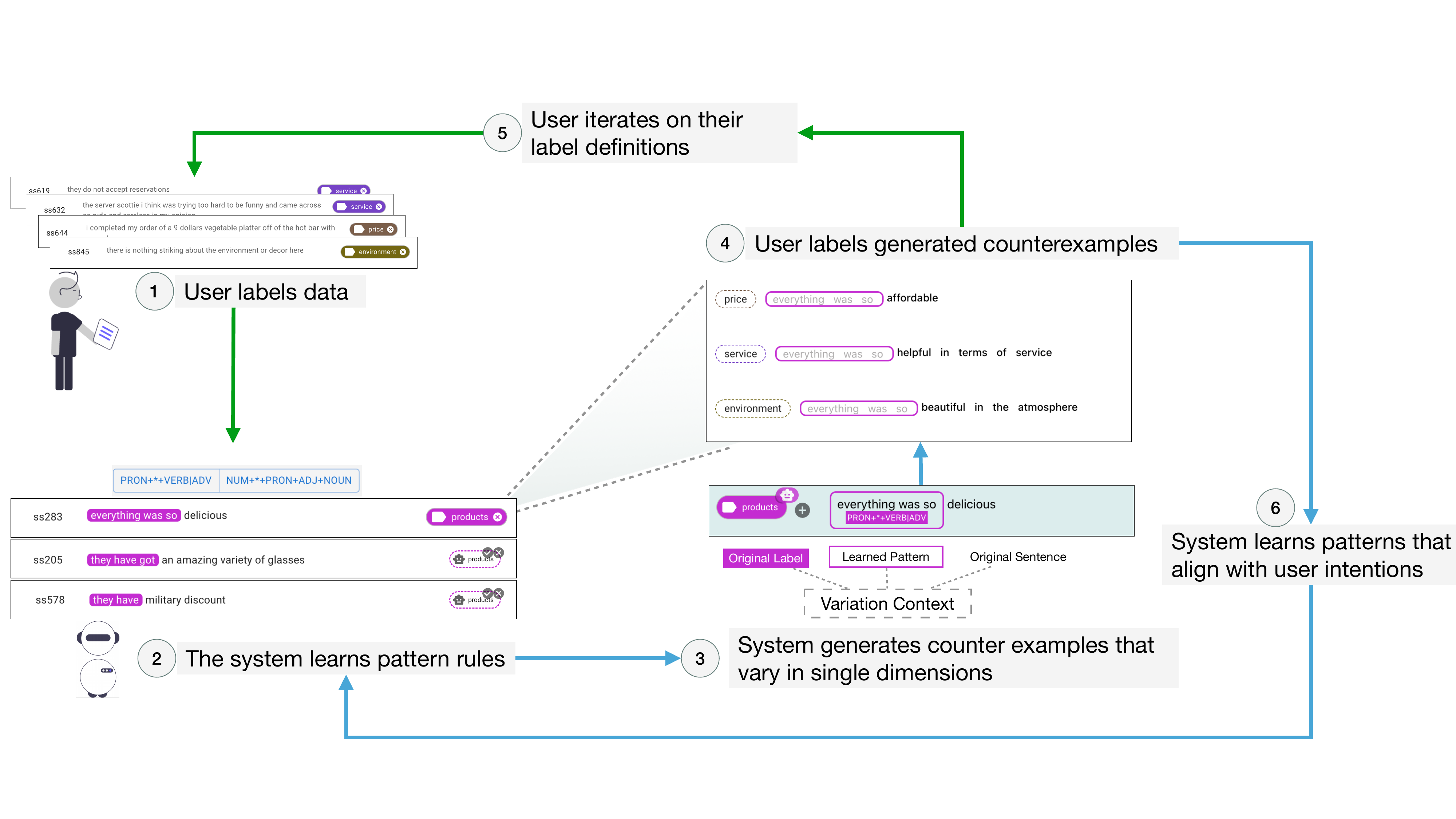}
    \Description{This figure illustrates an iterative process where a user collaborates with a system to refine label definitions. The user begins by labeling data, after which the system learns pattern rules based on these labels. The system then generates counterexamples by varying specific dimensions of the original sentences. The user reviews and labels these counterexamples, providing feedback to refine the system’s understanding. This process continues, with the user iterating on label definitions and the system learning patterns that increasingly align with the user's intentions.}
    \caption{\revision{A user is iteratively teaching a neuro-symbolic model to distinguish between different concepts (labels). The process begins with the user labeling some data points~(1). This allows the neuro-symbolic model to learn pattern rules about the label and suggest annotations to unseen data points~(2). As the user reads and accepts or rejects model-suggested labels, \toolname{} uses an LLM to generate counterfactual examples that structurally resemble the original data point and match the original patterns but have different predicted labels~(3). When presenting the generated counterfactual examples to the user \toolname{} emphasizes the changed parts and de-emphasizes the carried-over parts of each sentence while highlighting (in the associated label's color) where the current neuro-symbolic model would fail on the generated counterfactuals~(4). The user then assigns labels to the generated counterfactual examples by accepting or rejecting the LLM-generated labels to be used in consecutive model training~(5). As the user provides feedback through labeled data, the model iteratively learns and adjusts its decision boundary, to better align with the user's mental model and labeling criteria~(6).}}
    \label{fig:system-architecture}
\end{teaserfigure}

\maketitle

\input{01-intro}
\input{02-relatedwork}

\input{03-system}
\input{04-user-study}
\input{05-discussion}

\begin{acks}
This work was supported in part by an IBM Ph.D. Fellowship, an AnalytiXIN Faculty Fellowship, a Notre Dame-IBM Technology Ethics Lab Award, a Google Research Scholar Award, Alfred P. Sloan Foundation FG-2023-19960, and NSF Grants IIS-2107391, CCF-2123965, CMMI-2326378, and CNS-2426395. Any opinions, findings, or recommendations expressed here are those of the authors and do not necessarily reflect the views of the sponsors. 
\end{acks}

\balance

\bibliographystyle{ACM-Reference-Format}
\bibliography{main}

\newpage
\input{06-appendix}

\end{document}

%% file: 00-macro.tex
\usepackage{ifthen}
\provideboolean{revising}
\setboolean{revising}{false}

\newcommand{\revision}[1]{#1}
\newcommand{\delete}[1]{}

\newcommand{\sgcomment}[1]{}
\newcommand{\tlcomment}[1]{}

    \newcommand{\tc}[1]{}

%% file: 01-intro.tex
\section{Introduction}

In supervised and semi-supervised machine learning (ML) pipelines, labeled data is a vital component of training and validating models~\cite{monarch2021human}. \revision{Interactive ML~(IML) methods, like active learning~\cite{arora2007active}, continuously apply human feedback during model training to iteratively build and refine the model~\cite{hassan2023d, lewis1995sequential, margatina2021active}. A targeted approach in IML is machine teaching (MT)~\cite{taneja2022human}, an interactive framework that allows users to devise and select useful data for labeling, with the goal of teaching the model relevant features during training~\cite{fails2003interactive, brooks2015featureinsight}.} Through labeled data, human users ``teach'' an underlying concept to the model~\cite{ramos2020interactive}.
In a MT pipeline, humans act as experts on concepts with an explicit goal of creating ML models through a teacher-student interaction. The approach of MT heavily resembles the co-adaptivity of human-to-human teaching \revision{in which the teacher illustrates concepts, assesses the learner's progress and evolution of a concept~\cite{kulesza2014structured, suh2019anchorviz}, and iteratively revises their teaching approach}~\cite{konzett2020co, taneja2022human}.

\revision{Prior work in IML has incorporated human input into model training and refinement through iterative processes of labeling and reviewing~\cite{ramos2020interactive}. A prominent form of input is when humans provide labels for representative training examples, with the goal of adjusting model parameters~\cite{settles2009active}. When training samples are scarce, model performance heavily depends on the quality of available training examples~\cite{dodge2020fine}. However, relying exclusively on existing examples is not ideal for tasks requiring nuanced understanding of user intentions, as these examples often fail to represent diverse and edge-case scenarios~\cite{gomes2024finding}. While using synthetic data for active learning has promising results to mitigate data scarcity~\cite{quteineh2020textual}, much of this work prioritizes optimizing model performance, offering limited support for human learning and critical reflection---an essential component of MT. To support users in building accurate conceptual models during model teaching, \citet{gillies2016human} argue that interfaces should be reframed to account for human cognitive processes. Data labeling as a cognitive task---including defining a concept or determining how two similar objects may have different labels---requires both comparison and integration~\cite{wisniewski1999makes}.}

To address these needs, we introduce an interactive tool called \toolname. \toolname{} presents novel interaction mechanisms inspired by two human cognition theories. First, the Variation Theory of human concept learning~\cite{marton2014necessary} informed a new approach to generate synthetic counterfactual data for users to annotate. Secondly, the Structural Alignment Theory~\cite{gentner2001structural} guides the design of \toolname{}'s interface for presenting generated counterfactual examples in batch, which assists users in perceiving and comprehending the alignable differences between data items and annotating these data items in batch.

\textbf{Counterfactual Data Generation.} In the counterfactual data generation phase, once a user annotates a small initial dataset, \toolname{} employs a Variation Theory (VT)~\cite{marton2014necessary}-based pipeline to create synthetic data. \revision{VT posits that human learning occurs when learners experience variation across critical and superficial aspects of a concept---through exposure to contrasting examples that systematically vary along different critical and superficial feature dimensions.} Inspired by VT, our pipeline \revision{starts with the neuro-symbolic model's current (and potentially imperfect) learned pattern rules, which can be thought of as feature dimensions.} It then generates \revision{counterfactual data that are syntactically and semantically similar enough to an already-annotated datum that they would be given the same label by the neuro-symbolic model's pattern rule, but different enough that they would be given a different label by a standard pre-trained large language model.}
Therefore, the generated data poses the hypothetical question~\cite{feder2021causalm}: ``How should the model's prediction change if certain aspects of the input were altered?’’ 
\delete{This method offers several benefits. First, it generates annotations that are particularly valuable for model training, as they often lie close to the conceptual boundaries. These boundaries are critical because minor differences in interpretation can significantly affect the outcome. Annotating these data points helps align the model’s learning with the user's understanding. Second, this process encourages users to engage in analogical thinking, enhancing their reflection on and understanding of the underlying concepts. Finally, these matched patterns enable the presentation of alignable differences, a concept derived from structural alignment theory, which we will explain later.}  \looseness=-1

\revision{Consider this analogy {to illustrate the counterfactual approach for refining concept boundaries}; a user and a model are negotiating how to define a sandwich. Although both may start with their own definition, neither is accurate or specific. We suppose that the user starts with a definition \textit{``a sandwich is two slices of bread with meat in between.''} Although this definition may be a good candidate, it misses important features that make a sandwich a sandwich. In this analogy, our proposed approach would ask the user if grilled cheese is a sandwich as a counterfactual proposition. This counterfactual highlights the discrepancy between the outcome of executing the proposed rule (grilled cheese \textit{would not} be a sandwich because of the lack of meat) and a pre-trained model's understanding (grilled cheese IS a sandwich according to the LLM).  By highlighting the difference between the original definition (having meat in between) and a new synthesized definition with the new counterfactual example annotated~(having any filling in between), the two entities are able to iteratively negotiate and reach a shared definition of sandwiches.  This iterative redefinition would continue along different feature dimensions of a sandwich (like the number of breads, the type of bread, etc.) until an optimal shared agreement is reached. It is worth noting that this is a collaborative negotiation rather than a debate. Neither of the parties came to the interaction with a complete definition of sandwiches and tried to convince the other party. Instead, they came with an incomplete definition, were open to reflect on and change their own conceptual understanding based on interactions, and shared the goal of reaching clearer boundaries of the target concept.}

\delete{Our pipeline introduces several distinct differences compared to existing Active Learning (AL) and data augmentation methods. Similar to existing Active Learning (AL) approaches, our method focuses on presenting users with the most informative data points for annotation within a human-in-the-loop framework. However, conventional AL methods generally assume that the human's {concept understanding and} annotation behavior remain constant and prioritize model performance optimization. This leads to minimal consideration of the cognitive processes humans employ during annotation. In contrast, our pipeline is designed to consider and enhance these human cognitive processes, in addition to improving the model's learning. Compared to conventional data augmentation techniques used in machine learning, which primarily utilize rule-based methods to generate new data with predictable labels without requiring further human input, our approach differs significantly.}

\delete{Although it also aims to create additional synthetic data to complement existing datasets, our pipeline specifically generates data items {the model is uncertain about at a given moment}. This uncertainty necessitates continued human involvement in the annotation process, ensuring a dynamic interaction between the user and the model. This approach not only enriches the dataset but also aligns more closely with the evolving understanding of both the human annotators and the machine learning model.}

\textbf{Structure Aligned Data Rendering.} \toolname{} employs Structural Alignment Theory \revision{(SAT)}~\cite{gentner2001structural} to support the user's cognitive process of interpreting and understanding varying \revision{generated data}. \delete{through the visualization of alignable differences} According to SAT, \revision{humans compare two similar entities by trying to find structural alignments between them, and then comparing corresponding elements, with a special focus on differing aligned elements.} \delete{comparing structured representations of entities involves aligning them to highlight and project their differences and similarities.} \toolname{} \revision{contrasts the original user-labeled data with generated counterfactual examples, by visually emphasizing portions of the counterfactuals that have changed over the parts that stayed the same.} \delete{utilizes this approach to contrast the original data previously labeled by the user with the generated counterfactual examples. The tool highlights portions of the counterfactual data that correspond with the neuro-symbolic model's learned patterns, as illustrated in~(Fig~\ref{fig:system-architecture}).}\delete{ By visually highlighting the discrepancies between their own label definitions and the model's learned decision boundaries.} \delete{To help users focus on features that impact both the model's learning and their understanding, } Specifically, \toolname{} displays unchanged elements of the counterfactual examples in gray. In contrast, elements that have changed---and may thus influence a change in label---are highlighted in black, drawing the user's attention to these critical differences. \revision{By assisting users in comparing discrepancies between their own label definitions and the neuro-symbolic model's learned decision boundaries, users can provide annotated data that can update the model to align with their expectations.}

We validated the usability and effectiveness of \toolname{} in a lab study with 18 participants. Participants reported that the tool's workflow enhanced their understanding of both the underlying model's behavior and the data itself. \toolname{} was also shown to significantly improve annotation efficiency and \revision{improve the model's performance in learning user intents}. These findings point to important design implications for future human-AI alignment efforts. Specifically, they underscore the need for co-adaptive systems that can evolve along with users' mental models and definitions of labels. \revision{Our findings also highlight the implications of closing the loop in supporting human cognition with proposed interactive ML pipelines.} This adaptability is crucial for fostering deeper and more effective interactions between humans and artificial intelligence in complex data environments.


This paper makes the following contributions:
\begin{enumerate}
    \item We contribute a SAT-based rendering method for counterfactual examples.
    \item We built \toolname{} to understand how Variation Theory-based counterfactual generation~\cite{geebregiabher2024leveraging} combined with SAT-based counterfactual rendering affects the human's experience in co-adaptive machine teaching. 
    \item A lab study with 18 participants to demonstrate the usability of \toolname{} and its effectiveness in improving annotation efficiency, enhance the model's learning, and facilitating co-adaptive learning where users gain insight into the states of the model and reflect on their own understanding.
\end{enumerate}


%% file: 02-relatedwork.tex
\section{Related Work}

\subsection{Human-AI Alignment through Data Annotation}

Alignment is a bilateral process; it refers not only to AI acting according to human intentions but also to humans better leveraging AI by understanding the mechanisms behind it~\cite{shen2024Towards}. In this process, both the trainer and the learner aim to develop and maintain an accurate understanding about the target concept. In the machine learning pipeline, one way users show their intentions is through annotating data~\cite{zhang2016process}. However, when labeling under uncertainty or in the initial training phases, users may lack an understanding of the model capabilities.

Machine teaching, a part of the human-in-the-loop approach, has been used as a process in which a human expert (the ``teacher'') provides guidance to a machine learning model to help it learn important and robust features for decision making~\cite{Swati2023Human}. In this workflow, humans continuously guide the model to align its learning with their intentions. A common way to do this has been through data annotation~\cite{zhang2023peanut}. To align model training with human intent through data annotation (1) the human needs to understand the current state of the model and (2) the human should be able to take action to steer the model in their desired direction~\cite{zhang2016process}. This signifies that alignment demands not only that humans should be able to steer AI in their desired direction but also that humans need to understand the current state of AI and what it has learned in order to better utilize the latest AI advancements. 

\revision{Previous work shows that, although incomplete, users may possess some knowledge about target concepts~\cite{szymanski2024comparing,szymanski2024limitations}, which they use as a reference when building classifiers IML~\cite{suh2019anchorviz, chang2016alloy}. AnchorViz~\cite{suh2019anchorviz} and Alloy~\cite{chang2016alloy} provide more context to the users through clustered data to help them determine what should and should not belong to different labels. As users interact with more context and data, their definition of the concept could shift and evolve~\cite{kulesza2014structured}. A key challenge here lies in designing interactive systems that both acknowledge the dynamic nature of users’ conceptualizations and transparently illustrate how these evolving inputs influence the model’s outputs~\cite{ramos2020interactive}.} Consequently, this also requires the model to demonstrate how it adapts to the user's preferences to enhance its interpretability and increase human trust in the system~\cite{Jacovi2021Formalizing}. 








\subsection{Active Learning and Counterfactuals}

Active Learning~(AL) in machine learning is an approach in which the learning algorithm selectively chooses which data points should be labeled for training~\cite{felder2009active}. The primary goal of this approach is to minimize the amount of labeled data needed to learn a target concept by requesting labels, usually from humans, for the most informative examples, allowing the concept to be learned with fewer annotations~\cite{arora2007active}. To accomplish this, pool-based~\cite{zhang2022survey} and instance synthesis-based~(also called query synthesis)~\cite{schumann2019active} selection strategies have been used. An example synthesis-based strategy can be especially effective in domains with ambiguous and subjective labeling, as it creates new, potentially informative examples that broaden the distribution of labeled data while continuously adapting to the model. Counterfactual generation can be seen as a more targeted form of query synthesis. While query synthesis broadly creates new examples to inform the model, counterfactuals specifically explore ``what-if'' scenarios by modifying existing instances in meaningful ways~\cite{robeer2021generating}. Counterfactuals have been used to test the sensitivity of a model to small changes to refine its understanding of causal relationships~\cite{feder2021causalm}. Both query synthesis and counterfactuals aim to generate examples that are useful to the model. However, not all examples are equally informative for the model or equally easy to label for humans~\cite{culotta2005reducing}.

Due to the inherent complexity of language and its discrete nature, natural language counterfactual generation presents greater difficulties compared to structured and image data. For that reason, natural language counterfactuals have seen limited exploration. Previous works have proposed both generation-based and augmentation-based approaches. \citet{schumann2019active} uses variational auto-encoders to generate examples from uncertain regions in a model's latent space to improve a classified model. Alternatively, \citet{dixit2022core} uses a retrieve-then-edit framework to generate counterfactuals by conditioning on naturally occurring data. With the generative capabilities of large language models~(LLMs), there have been more efforts to automatically generate plausible counterfactuals by augmenting real examples~\cite{chen2022disco, wu2021polyjuice, dixit2022core}. Polyjuice~\cite{wu2021polyjuice} uses a fill-in-the-blank approach to generate counterfactuals by perturbing specific parts of a sentence according to predefined control codes (e.g., negation, quantifiers, or lexical modifications). Similarly, DISCO~\cite{chen2022disco} uses spans to determine what needs to change. Although both approaches use different methods to determine what needs to change, they find that the counterfactuals improve the downstream model's performance.

Most previous research in counterfactual generation has focused on the model side by either generating counterfactuals to improve the model's performance or explaining its behaviors post hoc.   
As intended with the design of \toolname{}, we believe this process should encourage users to engage in analogical thinking, enhancing their reflection and understanding of the underlying concepts when they are not well defined. \toolname{} uses human cognitive learning theories to support human annotation efforts for model training. Specifically, we use  Variation Theory of learning~\cite{marton2014necessary} which states that for learning to occur, some aspects that define the concept being learned must vary while others are held constant.

\subsection{Supporting Sensemaking Based on Variation Theory and/or Structural Alignment Theory}
Structural Alignment Theory~(SAT)~\cite{gentner2001structural} is a cognitive theory that explains how people make sense of concepts by comparing relational structures between two items. It states that understanding and sensemaking involve mapping the relationships between elements, especially in complex and ambiguous tasks. While SAT focuses on similarities and differences within alignable structures, \citet{estes2004importance} highlights the significant role of bringing salience to ``non-alignable'' differences. In decision making, SAT argues that people tend to focus on alignable differences---features that can be directly compared---rather than on differences that cannot be easily aligned. Although its application remains limited outside of the field of psychology, 
SAT has been used in broad domains such as consumer behavior research~\cite{kivijarvi2018advancing}, spatial data analysis~\cite{cruz2008structural}, Human-Robot Interaction (HRI)~\cite{booth2022revisiting}, and Human-Computer Interaction (HCI)~\cite{yan2022concept}. The last two prior works also combine Variation Theory (VT) and SAT together, as we did (i.e., a corollary of SAT referred to as Analogical Transfer/Learning Theory).

In developing ML models, annotators often engage in a process of comparing instances within the data, not just to match surface-level features such as keywords, but to discover relational patterns that inform their label definitions and boundaries~\cite{lee2012based}. Therefore, SAT provides a useful and applicable framework for thinking about data annotation, particularly in domains where annotators continuously define and refine labels during model training. 

Comparison as a means for sensemaking also finds relevance in modern tools designed for large-scale output sensemaking; two previous tools explicitly leverage SAT and Variation Theory (VT)-based designs. Positional Diction Clustering~(PDC)~\cite{gero2024supporting} is a structure mapping engine~\cite{gentner1997structure} introduced to facilitate sensemaking of many LLM responses to the same or similar prompts. It finds a structural mapping across all the LLM responses and can highlight alignable differences within that alignment using text salience.  ChainForge~\cite{arawjo2024chainforge} provides an interface to compare model outputs, where the variables that create the systematic variations in models and model prompts correspond to dimensions of variation in Variation Theory. Both systems enabled users to quickly identify variations and patterns within the data and support exploration and hypothesis testing. 

In line with previous work, \toolname{} aims to support a user's efforts in the disambiguation of concepts through structural comparisons of counterfactual data in the context of machine teaching. Specifically, \toolname{} highlights variations between data items to help users identify inconsistencies between their own label interpretations and the model predictions. In the context of interactive ML, where users are in charge of labeling data with the goal of influencing the model's training, interactive error correction is crucial~\cite{cabrera2023did, shrestha2023exploratory}. By presenting relational structures (e.g., causal chains for wrong predictions in counterfactuals) instead of just showing learned feature importance, \toolname{} helps users understand how the system makes decisions and identify how their annotations could change the model's behavior.

%% file: 03-system.tex
\section{System Description}

\revision{Machine teaching with exploratory data labeling requires users to provide distinguishing training examples of a concept~\cite{habibelahian2022exploratory}. When labeling similar data with subtle differences, users must compare data points while incorporating concept definitions~\cite{wisniewski1999makes}. In the context of co-adaptive learning, supporting the intertwined evolution of both the user's understanding and the model's learning is crucial~\cite{dudley2018review}.}

\revision{To support users in providing informative training examples based on the model's current state,} the back-end pipeline of \toolname{} integrates a neuro-symbolic approach with LLMs to guide the synthesis of counterfactuals that resonate with human cognitive processes. Building on methods proposed in PaTAT~\cite{geebregiabher2024leveraging}, \toolname{} first generates human-readable neuro-symbolic pattern rules from partially labeled text data for classification. Then, \toolname{} generates synthetic counterfactual text data that share syntactic and semantic patterns with the original text data, yet differ in the outcomes of the labels predicted by an LLM. This approach allows users to annotate data near conceptual boundaries, improving their understanding of the current limitations and capabilities of the neuro-symbolic model, as well as possibly facilitating the evolution and refinement of their own notions of the concepts involved. These annotations are pivotal for the model to learn user-specific values, preferences, and goals.

Counterfactual generation based on Variation Theory allows the implementation of Structural Alignment Theory on \toolname’s interface design to render and highlight alignable differences between original and generated data items in real time. This facilitates the user's sense-making process of discerning key differences among items and reasoning about how and why they lead to different labels during annotation. Thus, ultimately improving the efficiency and effectiveness of the model teaching process.

The section begins by outlining the design goals of \toolname, presents a motivating user scenario, discusses its key features, their rationale, and how they connect to the design goals, and concludes with implementation details.

\subsection{Design Goals}
\label{sec:design_goals}
\delete{To support users in building accurate conceptual models during model teaching, \citet{gillies2016human} argue that interfaces should be reframed to account for human cognitive processes. Data labeling as a cognitive task -- including defining a concept or determining how two similar objects may have different labels -- requires both comparison and integration~\cite{wisniewski1999makes}. In the context of co-adaptive learning, supporting the intertwined evolution of both the user's understanding and the model's learning is crucial.}
\toolname{} is designed to assist the user in their data annotation efforts and sensemaking while simultaneously providing useful training data for the model's learning. Specifically, \toolname{} follows these design goals:




\paragraph{\textbf{DG1: Facilitating User Understanding of the Model's Current State}} 
\revision{Teaching} is inherently exploratory. Aligning model training \revision{through MT} with human values iteratively requires clear user understanding of the model's current state \cite{shrestha2023exploratory}. This is achieved by providing transparency in the model's decision-making processes, emphasizing areas of uncertainty or salience, and illustrating the model's data interpretations. Such insights enable users to recognize where the model excels, where it falters, and how it evolves over time, empowering them to make informed decisions to refine and enhance the model through targeted annotations.


\paragraph{\textbf{DG2: Augmented Data Should Refine the Model's Decision Boundary }}
The impact of labeled data is crucial in refining the model’s decision boundary, particularly within areas of high uncertainty~\cite{yang2016active}. Therefore, the data annotated in each iteration should improve the model's understanding of its decision boundaries. Augmented data \delete{must not only increase the volume of labeled training data but} should spotlight edge cases and ambiguous instances near the decision boundary~\cite{bunzel2024identifying}. Addressing these cases improves the model's generalization capabilities and offers users deeper insights into its decision-making processes.


\paragraph{\textbf{DG3: Enhancing Interface Support for User Annotation of Generated Data}}
Research has identified data annotation as a critical bottleneck in the model training pipeline~\cite{aroyo2022data}. While the original data is a fixed resource, augmenting data can be strategically aligned with users' cognitive processes to facilitate sensemaking. Traditional annotation methods, which rely heavily on manual reviews, often fall short in handling complex datasets and counterfactual examples. To address this, \toolname's initial phases of data integration employ interaction mechanisms that simplify the identification and annotation of data points. These mechanisms contribute to a more diverse and comprehensive dataset and clarify the intricacies of decision boundaries. This requires the development of interfaces and visualizations  that demystify the generated data, allowing systematic variation and coverage across the concept space.


\subsection{Motivating User Scenario}
\label{sec:user_scenario}

\begin{figure*}
  \centering
  \includegraphics[width=\linewidth]{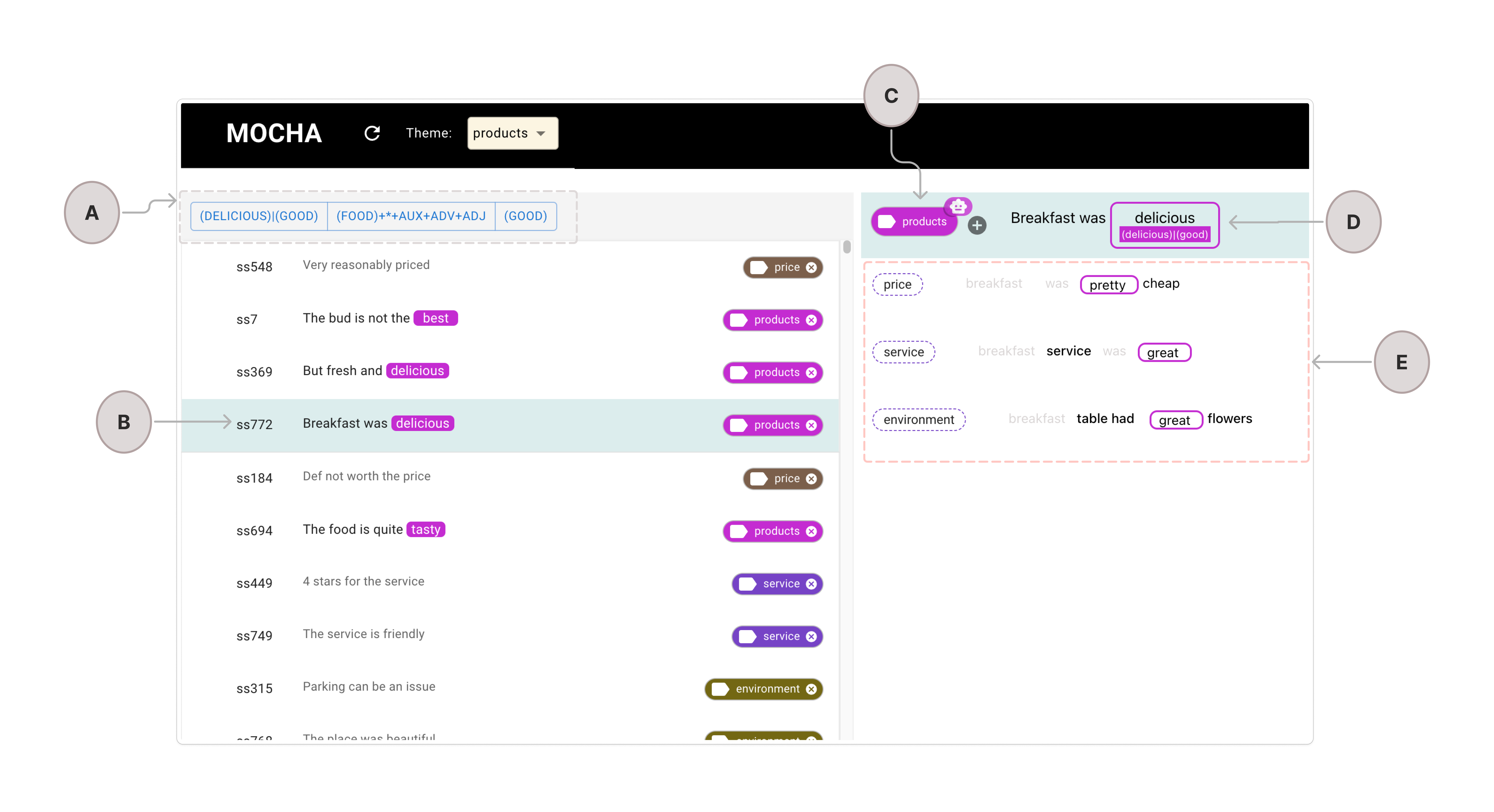}
  \Description{This figure shows an interactive labeling model training and data labeling interface. At the top (A) displays the model's learned pattern related to "delicious" and "good" for the label product. A list of labeled sentences (B) is shown, with one sentence, "Breakfast was delicious" (C), highlighted and tagged under the "products" label. The word "delicious" is emphasized (D) as part of the pattern matching. On the right, suggestions (E) offer alternative sentence variations across different dimensions (e.g., price, service, environment) to assist users in refining label definitions.}
  \caption{\toolname{} uses neuro-symbolic pattern rules~(A) to generate counterfactuals. For each example labeled by the rules~(B), \toolname{} generates counterfactual examples that match the original patterns~(D) but belong to a other than the original label~(C). The generated counterfactuals are then rendered below the original example with highlighting of what has changed and what has stayed the same~(E) for each alternative label.}
  \label{fig:all_patterns_view}
\end{figure*}

\begin{figure}
    \centering
    \includegraphics[width=\linewidth]{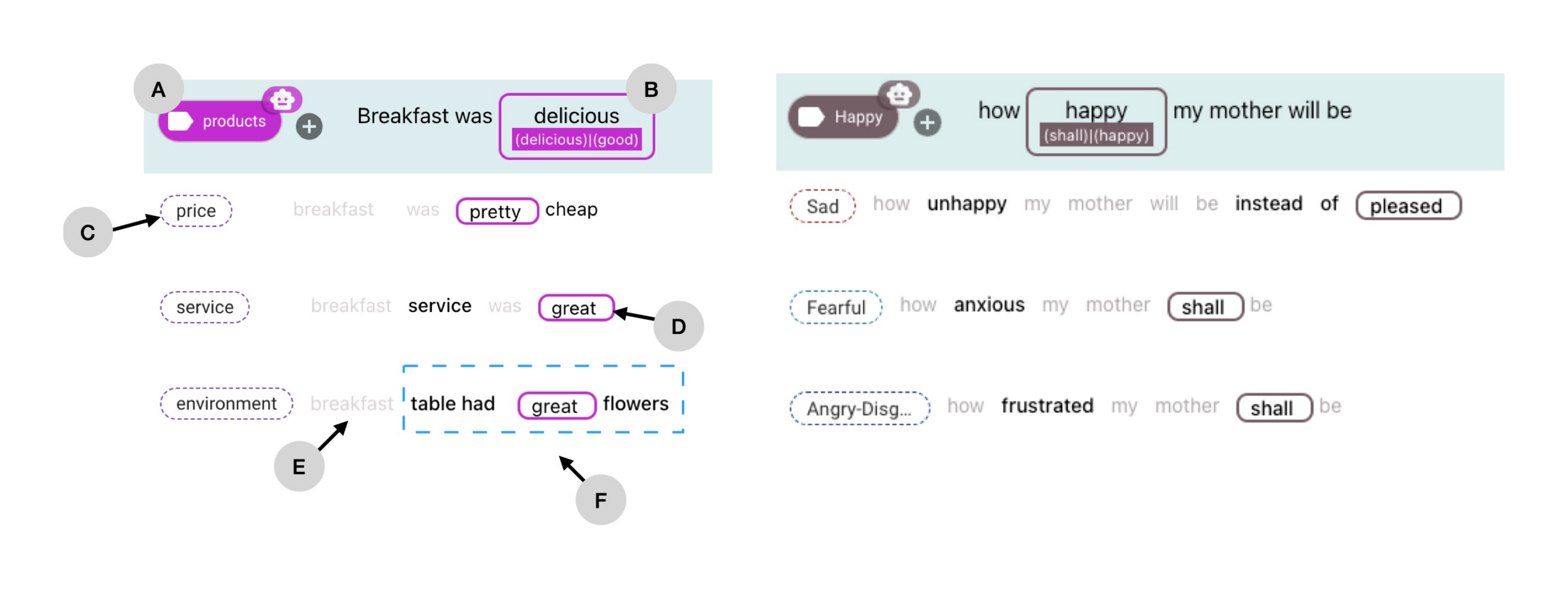}  
    \Description{This figure presents two examples of counterfactual variations based on user-labeled sentences. On the left, the sentence `Breakfast was delicious' (A) is labeled as related to "products' and the word `delicious' is highlighted (B). Suggested counterfactuals are displayed for `price' with `pretty cheap' (C), `service' with `service was great' (D), and `environment' with `table had great flowers' (E), showing variations in different contexts (F). On the right, the sentence `how happy my mother will be' is labeled as `happy' and suggestions such as `unhappy,' `anxious,' and `frustrated" replace `happy' to reflect emotional variation in other contexts.}
    \caption{\toolname{} facilitates analogical reasoning using visual cues. For each model-labeled example (A) and its corresponding learned neuro-symbolic rule (B), counterfactual examples are generated for a set of target labels (C). Phrases consistent with the original example are displayed in gray text (E), while varying phrases are displayed in black text for visual salience (F). Additionally, the text of the counterfactual that would mislead the neuro-symbolic model into classifying it as the original label (by matching the original label's rule) are highlighted in the theme color (D), helping users understand how their annotations contribute to model updates.}
    \label{fig:counter-zoom}
\end{figure}

Alice is working on training a model to classify text snippets from online reviews. Although she has an idea of the possible labels in the dataset, there is some uncertainty about which features are most relevant for the model’s training and how these features define the feature space for each label. \revision{She is also exploring how to differentiate between seemingly overlapping concepts, for example weather friendliness of a restaurant staff should belong to the label \textit{service} or \textit{environment}.  Alice decides to use \toolname{} to iteratively label data and train a classifier model.
While she iteratively trains the model she expects to have a more complete understanding of what appropriate features each label should contain and have a solid definition of the labels.} \delete{Alice decides to iteratively teach the model to ensure that it learns the appropriate features to define each concept. Her goals are twofold: (1) to train a model that effectively discriminates between different labels, and in the process (2) understand and solidify her understanding of each label's definition.}

Alice uploads her dataset into \toolname, \revision{a tool designed to iteratively align the model’s learning process with the user’s mental model. She begins the process of assigning labels to data items. After every ten annotations, Alice notices the tool suggests labels to unlabeled data items based on learned pattern rules~(Fig~\ref{fig:all_patterns_view}-A). These rules, generated by a neuro-symbolic model, are constructed using program synthesis to find an optimal combination of domain-specific language that best fit Alice's labeled positive and negative examples~(details in Appendix~\ref{app:patat-rules}). The rules aim to capture the discerning features of each label based on Alice’s annotations.} While some of the suggested rules align with Alice’s mental model, others may be too broad or fail to accurately reflect her intended label definitions.

\delete{During her labeling, Alice notices that \toolname{} has learned a rule that while valid is too broad.}To see data with suggested labels~(Fig~\ref{fig:all_patterns_view}-C), Alice can click on a pattern rule~(Fig~\ref{fig:all_patterns_view}-A) to filter data points that match the selected pattern rule~(\textit{DG1}). \revision{As Alice clicks on the data point to assign a label to it,} \toolname{} generates counterfactual examples~(Fig~\ref{fig:all_patterns_view}-{E}) \revision{that are structurally close to the original data point she is currently labeling}. The generated counterfactual examples match the learned neuro-symbolic rules~(Fig~\ref{fig:counter-zoom}-{D}) but are labeled differently by an off-the-shelf LLM~(Fig~\ref{fig:counter-zoom}-C, \textit{DG2}).

After labeling the original data point~(Fig~\ref{fig:all_patterns_view}-B), Alice moves on to label the generated counterfactual examples. When examining them in batch, her attention is drawn to the differences between the original and counterfactual examples, with the differing parts highlighted in \textbf{black} and the unchanged parts in \textit{gray}~(Fig~\ref{fig:counter-zoom}-E and F, \textit{DG3}). \revision{Each of the generated counterfactuals match the learned patterns and are incorrectly labeled by the neuro-symbolic model. With the generated counterfactual examples, Alice sees what the model is learning and failing to learn. By labeling the counterfactual examples, Alice provides the model with additional data points to fortify the interpretation of a label, as shown by the learned pattern rules. At the same time, making labeling decisions on data points that are at the model's decision boundary helps Alice refine or confirm her own understanding of the label.} The labeled counterfactual examples are then used during consecutive model training. \delete{After making some annotations Alice notices the generated rules reflecting the model's learning update~(Fig~\ref{fig:all_patterns_view}-A).} After each round of annotation the neuro-symbolic model updates its learned rules \revision{and this process continues until Alice's  interpretation of each label finally aligns with the model's.}


\subsection{Key Design Features}
This section describes the following key design features of \toolname{} to support the continuous training and alignment between humans and AI.

\subsubsection{The Generation of Alignable Counterfactuals}
\label{sec:alignable_counter_generation}
Counterfactual examples enhance the training of a model by generating misclassified instances, which the model can then correct through retraining. This process works because counterfactuals reveal edge cases in the model's learned decision boundaries. \citet{geebregiabher2024leveraging} argued that counterfactual generation that follows the principles of VT allowed the introduction of discriminatory variance for the model to learn on. According to Variation Theory, learners better understand concepts by observing variations along critical features (dimensions of variation) that define that concept and, separately, observing variations along superficial features that do not define that concept---all while other features, when possible, are held constant. For instance, students who see triangles in different orientations can deduce that the defining characteristic of a triangle is its three sides and not its orientation. Building on that, in \toolname, we adopt this approach to develop a user-facing interface that supports a user's learning in parallel to the model's learning. 

To integrate VT into counterfactual generation, this method begins by identifying key features and introducing variations while still satisfying the predefined neuro-symbolic pattern rules~\cite{gebreegziabher2023patat} that currently define the machine-learned concept; these pattern rules are interpretable features the model has learned as critical for a given concept~\revision{(see details in Appendix~\ref{app:patat-rules})}. 
The rules encompass lexical, syntactic, and semantic elements---including part-of-speech tags, word stems, synonyms (soft matches), and entity types---organized in regex-like patterns.  These patterns help capture commonalities across datasets with similar labels~(Fig~\ref{fig:all_patterns_view}-A).

In other words, for each label, the learned neuro-symbolic patterns reflect the model's current interpretation of that label; data points that match the model's linear combination of the neuro-symbolic patterns would be classified with that label. Therefore, when the learned patterns are inaccurate, the generated counterfactuals should provide examples that match the sub-optimal pattern but likely correspond to a different label. 
To facilitate generating counterfactuals along this dimension of variation, our approach starts by prompting an LLM to generate candidate phrases that match the learned pattern~(Appendix~\ref{sec:candidate-phrases}). For example, for a data item that was labeled ``product'', the sentence \textit{``Breakfast was \textbf{delicious}''} that matches the pattern \textit{`(delicious)|(good)'} will have phrases \textit{[`well priced', `pretty cheap', `worst deal', `good but overpriced']} generated as candidate phrases for a target label ``price''. The candidate phrases are used to enforce that the augmented counterfactual always includes the learned neuro-symbolic pattern. In the counterfactual generation prompt~(Appendix~\ref{sec:counter-generation}), we explicitly instruct the LLM to modify the original example, making minimal changes while still including the generated candidate phrases, to change the original label into a set of different target labels~(see Algorithm~\ref{alg:counter-generation})~(DG2).

\begin{algorithm}
    \caption{Generates counterfactual data based on learned neuro-symbolic patterns}
    \label{alg:counter-generation}
    \begin{algorithmic}[1]
    \Require Original dataset $D$, User label $UL$ ,Target label $L$
    \Function{GenerateCounterfactuals}{$D, UL, L$}
        \State Initialize $D_{cf}$ as an empty dataset
        \For{each $d \in D$}
            \State $p_{d} \gets \text{GetSymbolicPattern}(d, UL)$ \Comment{Generate patterns based on the user assigned labels}
            \State $candidatesPhrases \gets \text{GenerateCandidatePhrases}(d, p_{d}, L)$\Comment{Generate phrases that match the pattern but are about target label $L$}
            \State $variations \gets \text{GenerateVariations}(d, candidatesPhrases, L)$ \Comment{Change parts of the sentence with one of the candidate phrases}
            \For{each $v \in variations$}
                \State $cf \gets \text{CheckPattern}(v, P)$ \Comment{Check if counterfactual matches pattern $P$}
                \If{$cf$ successfully flips the label to $L$}
                    \State Add $cf$ to $D_{cf}$
                \EndIf
            \EndFor
        \EndFor
        \State \Return $D_{cf}$
    \EndFunction
    \end{algorithmic}
\end{algorithm}

\subsubsection{Facilitating User’s Perception of Similarity and Differences}

Assisting users to perceive consistencies and variations across different data points enables them to quickly develop an accurate mental model~\cite{noe2022learning, gero2024supporting}. To support this process, \toolname{} draws design inspiration from the Structural Alignment Theory~\cite{gentner2001structural} of how humans compare and contrast objects. Structural Alignment Theory states that humans naturally look for structural mapping between representations of objects to help them understand, compare, and infer relationships between said objects. In our context, these objects are the original examples and their counterfactuals. In rendering the generated counterfactuals, \toolname{} facilitates analogical reasoning through the mapping of the counterfactual label~(Fig~\ref{fig:counter-zoom}-C) and the generated counterfactual. 

To assist users in assessing the appropriateness of the counterfactual label for the generated example, \toolname{} uses visual cues enabled by the structure of variation induced by the Variation Theory-based counterfactual generation method in the previous step. Specifically, the changes introduced to change the original example into the counterfactual are highlighted in black~(Fig~\ref{fig:counter-zoom}-F) to draw user attention to them. This black text stands out more prominently than the unchanged text, which is rendered in gray~(Fig~\ref{fig:counter-zoom}-E). 

To determine this mapping, \toolname{} adopts the Levenshtein distance algorithm~\cite{zhang2017research},  which calculates the minimum number of edit operations at the word level required to transform the original example into the counterfactual example generated. Specifically, we define two types of edit operations: additions (inserting words) and deletions (removing words). The algorithm splits each sentence into its component words and identifies the shortest sequence of operations to transition from one sentence to another. Our objective is to minimize the number of operations while striving to maintain the longest continuous phrase unchanged between the two sentences. For example, given the original sentence \textit{``Breakfast was delicious''}, and a counterfactual sentence \textit{``Breakfast was pretty cheap''}, the algorithm would identify a delete operation for the word `delicious' followed by an insert operation for `pretty cheap'.

\subsubsection{Comparison Through Carried Over Matched Neuro-symbolic Rule}

\toolname{} aims to facilitate the user's understanding of where and why the neuro-symbolic model's understanding diverges from their expectations. To support this, it leverages the executable nature of the learned neuro-symbolic rules. \toolname{} highlights phrases in generated counterfactual that match the learned `imperfect' neuro-symbolic~(Fig~\ref{fig:counter-zoom}-D). This process can be understood as the common relational structure between the original and counterfactual examples. 

A key visual aid in this process is the use of theme colors~(Fig~\ref{fig:counter-zoom}-D), which highlight parts of the counterfactual that could have misled the model into making incorrect classifications. By applying a consistent and striking color to these terms, the system visually projects the model reasoning process onto the interface, making the inference projection process possibly easier to understand for users. From a cognitive perspective, the theme color aligns with the human's (theorized) structural mapping engine~\cite{gentner2001structural} by making relational discrepancies between the original and counterfactual examples more explicit. The model's reasoning is ``projected'' onto the counterfactual, enabling users to easily see which aspects of the counterfactual match the model’s existing rules, and which aspects lead to erroneous inferences~(DG1). This immediate feedback supports users in correcting the behavior of the model by adjusting the labels and refining the classification boundaries through targeted interaction.


\subsection{Implementation}

The interactive Web application of \toolname{} was developed using React\footnote{https://reactjs.org/}. The backend server utilized Python's FastAPI\footnote{https://fastapi.tiangolo.com/} framework to facilitate communication with OpenAI's API and to the frontend. In the backend, candidate phrases and counterfactuals were generated using the GPT-4o model from OpenAI through API calls. We used Firebase\footnote{https://firebase.google.com/} to track and store participant's interaction log data. Both the front-end and the back-end were hosted on a Google Cloud\footnote{https://cloud.google.com/} server for the user study.

%% file: 04-user-study.tex
\section{User Study}

The user study\footnote{The study protocol was reviewed and approved by the IRB at the lead author’s institution, where the study was conducted.} aims to evaluate the effectiveness of \toolname's key features, informed by Variation Theory~(VT) and Structural Alignment, on augmented data annotation and bi-directional alignment. Specifically, we evaluate the efficacy of the augmented counterfactuals on the user's annotation efforts, the model's learning, and the user's learning about the data and the relevant concepts.

The study specifically investigates the following research questions:
\begin{itemize}

    \item \textbf{RQ1:} Can Structural Alignment-based counterfactual rendering improve efficiency and lower cognitive load during the annotation process?

    \item \textbf{RQ2:} How useful are Variation Theory-based counterfactual generation in allowing the model to learn about the user's intents?

    \item \textbf{RQ3:} How useful are Variation Theory-based counterfactual generation and Structural Alignment Theory-based counterfactual rendering in allowing the users to learn about the data and clarify their intents to the model?
    
\end{itemize}

\subsection{Participants}
We recruited 18 participants (11 male, 7 female) with varying levels of experience in developing and training ML models, as detailed in Table~\ref{tab:participant-data} in Appendix~\ref{sec:study_participant_data}. The participants' ages ranged from 18 to 34 years. Among them, 4 participants self-reported having a beginner level with ML, 10 were intermediate, and 3 were experts. One participant had no previous experience with ML. More information on the demographics and backgrounds of the participants can be found in the Appendix~\ref{sec:study_participant_data}.

\subsection{Study Procedure}

Each study session lasted approximately 90 minutes and was conducted either in-person in a usability lab or virtually via Zoom (3 in-person, 15 virtual). After obtaining informed consent, the participants received a 5-minute tutorial on the key features of the tool. \revision{Participants were asked to train a multi-class classifier model by assigning one or more labels to data-items from the given list of labels. While participants worked with the same set of labels, they were told to follow their own interpretations of both the labels and data points.} The study used a think-aloud method, asking participants to verbally express their thoughts while annotating the data. The participants were then assigned one of two datasets and engaged with \toolname{} in three 25-minute sessions under different conditions. Here, they observed the neuro-symbolic model retrain and update following each annotation round. The sequence of these conditions was randomized, and details about the order and specific datasets are available in Table~\ref{tab:task-details}
 in Appendix~\ref{sec:study_task_details}. After each condition, participants completed a NASA-TLX~\cite{hart2006nasa} survey to assess their cognitive load. The study ended with a post-study questionnaire on \toolname's perceived usefulness and usability, and a semi-structured interview exploring the use of tools and experiences of participants in different conditions.

The study aimed to observe the participants' abilities to refine their subjective definitions of labels and the model's effectiveness in learning from the labeled counterfactual data. Initially, we presented participants with a dataset that included a predefined set of labels, allowing them the flexibility to define and redefine these labels as they annotated examples.

\subsubsection{Datasets}

In this study, we selected two datasets that are open to subjective interpretation and do not necessitate domain-specific knowledge from participants. Each participant worked with only one dataset throughout all conditions. The sampling method for each condition was designed so that no participant would encounter the same data item more than once across different conditions.

\begin{itemize}
    \item \textbf{Emotions}~\cite{alm2008affect} - Each entry in this dataset consists of a text segment extracted from tales with labels that indicate the predominant emotion conveyed. Participants worked with 5 categories---fearful, sad, happy, surprised, and anger-disgust. \revision{This dataset contained 100 independent samples for annotating and testing for each condition.}
    
    \item \textbf{YELP}~\cite{asghar2016yelp} - This dataset consists of user reviews of businesses (e.g., restaurants, retail stores) collected from Yelp with 4 categories---price, service, environment, and products. \revision{This dataset contained 160 independent samples for annotating and testing for each condition.}
    
\end{itemize}

\subsubsection{Conditions}

The study used a within-subjects design, where each participant experienced the system under three different conditions. To minimize carryover effects, the order of the conditions was varied for each participant.

\begin{itemize}
    \item \textbf{Condition 1 (Non VT Counterfactuals)}: Participants were asked to label counter examples generated without the use of VT and neuro-symbolic patterns. The generated counter examples were displayed without highlighting alignable differences for users to label.
    
    \item \textbf{Condition 2 (VT Counterfactuals without alignment)}: Participants were asked to label counter examples generated using VT with neuro-symbolic patterns but without highlighting alignable differences.
    
    \item \textbf{Condition 3 (VT Counterfactuals with alignment)}: Participants were asked to label counter examples generated using the same pipeline as \textit{Condition 2} with alignable differences highlighted in the interface.

\end{itemize}

\subsection{Study Results}

\subsubsection{Data Analysis}
\begin{figure*}[h]
    \centering
    \includegraphics[width=0.7\linewidth]{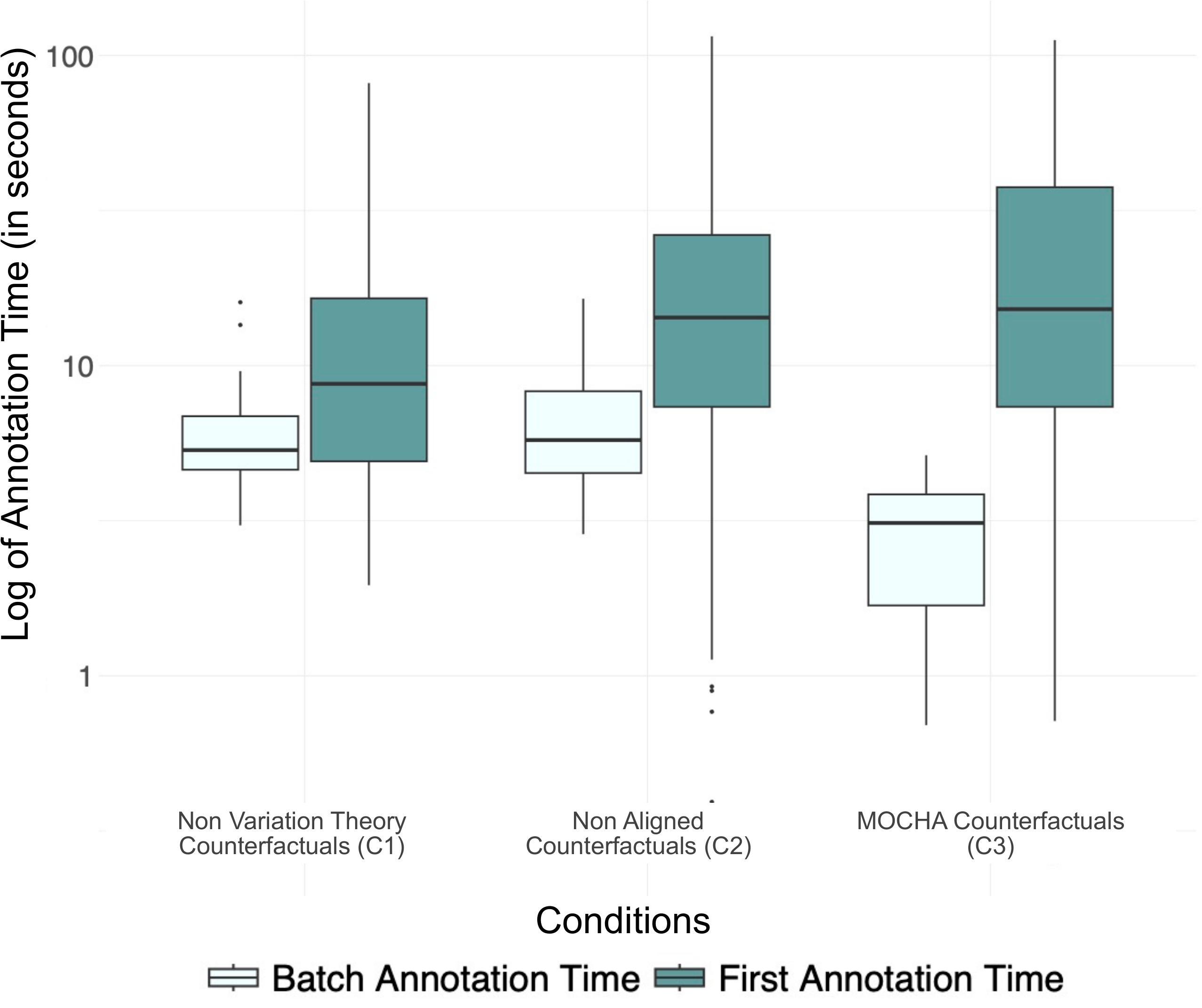}
    \Description{A box plot comparing the log of annotation time (in seconds) across three conditions: Non-Variation Theory Counterfactuals (C1), Non-Aligned Counterfactuals (C2), and MOCHA Counterfactuals (C3). The annotation time is divided into two categories: Batch Annotation Time and First Annotation Time. The y-axis represents the log of annotation time, with a range from 1 to 100 seconds. Each condition displays variability in annotation time, with C2 showing the widest spread in First Annotation Time, and C3 having the shortest median Batch Annotation Time.}
    \caption{Comparison of average annotation times under the three conditions. The table shows that while VT-based counterfactuals increase the time for the first annotation, SAT-based rendering significantly reduces the time for annotating each data point in the batch.}
    \label{fig:annotation-time}
\end{figure*}

We conducted statistical tests to compare responses to all the Likert scale survey questions, as well as the time spent annotating counterfactuals for each condition. 

To analyze the annotation efficiency, we first conducted a Kruskal-Wallis rank sum test~\cite{kruskal1952use} to determine if there were statistically significant differences in annotation time across the three conditions, because our data violated the homogeneity of variances assumption, making non-parametric methods more appropriate. To compare each condition against each other, we conducted a post-hoc pairwise test using the Wilcoxon rank-sum test~\cite{neuhauser2011wilcoxon} with continuity correction and Bonferroni adjustment. 

To analyze the Likert scale ratings of the participants from the post-study questionnaire, we first performed a Friedman test~\cite{friedman1937use} to determine whether there were statistically significant differences in participant ratings across the three conditions. Following this, we used Wilcoxon signed-rank tests with Bonferroni correction~\cite{neuhauser2011wilcoxon} and Kendall's W~\cite{field2005kendall} for post-hoc pairwise comparisons. These tests are nonparametric, which is appropriate for the ordinal nature of Likert scale ratings.

The analysis of the interview results was done through open coding~\cite{williams2019art}, in which two members of the team  coded the interview transcripts independently and then came together to consolidate.

\subsubsection{\textbf{RQ1:} Impacts of Structural Alignment interfaces on annotation efficiency and cognitive load.}

\paragraph{Improved annotation efficiency} 
To understand the impact of alignable differences on the participant's annotation efficiency with the generated text, we compare the average time it took for participants to read and make their first annotation on a generated sentence against the time it took them to complete a batch of annotations (i.e., annotating all generated examples associated with an original sentence as seen in Fig~\ref{fig:counter-zoom}). We calculate time of annotation as the difference between when the generated counterfactuals are rendered on the screen to when the participant assigned the first label, and subsequently all consecutive labels in that view. 

\begin{figure*}[h]
    \centering
    \includegraphics[width=\linewidth]{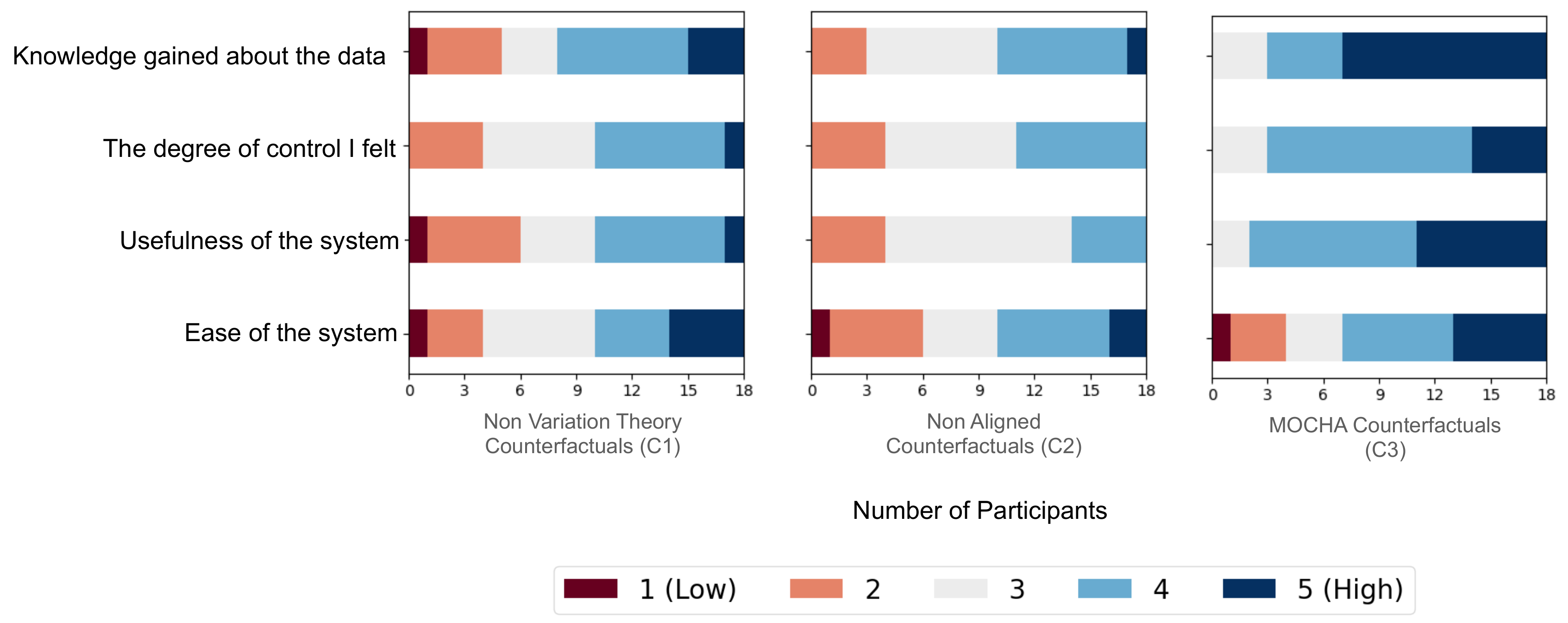}
    \Description{A stacked bar chart showing participants' ratings across four categories: Knowledge gained about the data, The degree of control I felt, Usefulness of the system, and Ease of the system for three conditions. The chart shows variation across conditions, with C3 generally receiving higher ratings, especially in the `Knowledge gained' and `Ease of the system' categories.}
    \caption{Participants' response to post study questionnaire comparing the three conditions.}
    \label{fig:post-condition-results}
\end{figure*}

\revision{On average participants annotated more data points in condition C3~(84.6, SD=26.0) compared to both C1~(70.82, SD=31.2) and C2~(69.65, SD=31.0).} As shown in Figure~\ref{fig:annotation-time}, participants spent a similar amount of time to annotate the first data point they saw across all three conditions. However, the batch annotation times were notably shorter in \toolname's counterfactuals (C3). The batch annotation times for counterfactuals with alignable differences (C3) were significantly lower compared to both counterfactuals generated without Variation Theory (C1) (p<.0001) and those generated with Variation Theory but without highlighted alignable differences (C2) (p<.0001). There were no statistically significant differences between C1 and C2 in their batch annotation times. This suggests that the efficiency improvements are primarily attributable to the specific features of C3, which is the rendering of alignable differences between the generated counterfactuals and the original example. Despite the lack of statistical difference between C1 and C2, the application of Variation Theory in C2 played a crucial role, enabling the introduction of alignable differences in C3.

\paragraph{Participants Perceived \toolname{} as More Useful in Condition 3}
In the post-condition questionnaire, participants compared their experiences across three different conditions. Figure~\ref{fig:post-condition-results} shows their subjective assessments in four dimensions: knowledge gained about the data, the degree of control they felt in changing the behavior of the model, the usefulness of the system, and the ease of using the system. C3 emerged as notably the superior condition, with participants reporting the greatest gains in knowledge about the data, increased control, and enhanced usefulness of the system compared to the other conditions.

Contrary to the reduced batch annotation time, we did not find a statistically significant difference in participant ratings of their perceived ease of use across the three conditions.

However, we find statistically significant advantage of C3 in the other three measures~(p<.05). The participant's rating of the knowledge gained in the three conditions revealed a statistically significant difference~(X\textsuperscript{2}(2, N=18)=11.6, p=.002) compared to the other conditions. The effect size, measured by Kendall's W, was 0.323, suggesting a moderate level of agreement among the rankings. The post-hoc pairwise test showed that the participants rated their level of knowledge gained about the data in C3 significantly different from both C1 (p=.02) and C2 (p=.007), while there was no significant difference between C1 and C2. This indicates that C3's structurally aligned rendering led participants to feel they gained significantly higher level of knowledge compared to the other conditions. These results align with the participant's usefulness ratings for the three conditions~(p=.0009) with an effect size of 0.39 suggesting a moderate to strong level of agreement among the ranking. Similar to the knowledge gained ratings, we find no significant differences between C1 and C2 but significant difference between C3 and both C1~(p=.01) and C2~(p=.002).

\paragraph{No difference in cognitive load observed}
 The NASA-TLX scores revealed no statistically significant differences in cognitive load between conditions (see Figure~\ref{fig:nasa-tlx} in the Appendix~\ref{sec:nasa-tlx}). This suggests that the variations in the counterfactuals generated and their presentation in C3 did not impose additional mental demands on the participants compared to C1. More research is needed to determine whether the interventions introduced in C3 could reduce cognitive load over extended use.


\subsubsection{\textbf{RQ2:} Effect of Variation Theory based counterfactuals on the model's learning}

A major goal of data augmentation is to improve the generalizability of the model~\cite{dixit2022core}. To that end, the annotated data should contribute a meaningful distribution to the model's training dataset \revision{to align with the user's own intentions}. To assess the downstream impact of the generated data on model training, we evaluated the model's performance when trained only on the real data and compared it to the model's performance when trained on the combination of the real data and the labeled counterfactuals. The participants' dataset, including the counterfactuals, provided a basis for comparing model performance under two conditions: with and without counterfactuals. Given that participants could flexibly define their own interpretations of the data, we use each participant's final labels as the ground truth. We calculated the precision and recall of the model using the finalized labeled data. 

\revision{In our experiment, we observed that the participants had moderate agreement with each other, measured by Fleiss' Kappa~\cite{fleiss1969large} (Yelp=0.67, Emotions=0.41). This relatively low agreement score for the emotions dataset suggests that the participants did not reach a strong agreement among themselves about how to label the data indicating  inherent ambiguity and subjectivity in the task. However, when the model is trained on these somewhat inconsistent labels and evaluated for agreement with their corresponding participant individually (using Cohen's Kappa~\cite{mchugh2012interrater}), the averaged agreement between the participant and their final model is stronger (Yelp=0.76, Emotions=0.87).}

\input{04-model_preformance_tab}

Table~\ref{tab:mode-results} presents a comparison of the final model performance for each participant. In most cases, the inclusion of labeled counterfactuals contributed to increases in both precision and recall. The inclusion of counterfactuals often resulted in a substantial increase in precision, indicating that the models were better able to correctly classify relevant instances while reducing false positives. This improvement suggests that the counterfactuals provided essential information that helped refine the models' decision boundaries. 

In scenarios where real data lacked annotated labels, the labeled counterfactual examples were instrumental in initiating the model's learning process and enabling the generation of relevant neuro-symbolic rules. \revision{For example, P4 observed the model learned pattern rules for the label \textit{`service'} from labeled counterfactuals generated for the label \textit{`price'}. Initially, there was insufficient annotated data for the model to generate patterns for the label \textit{`service,'} but with the inclusion of annotated counterfactuals, the model successfully learned a neuro-symbolic rule for the label \textit{`service'} which was \textit{(friendly)+*+NOUN} \footnote{This rule matches any sentence that has synonyms of the word friendly followed by a single wildcard and a noun}. Similarly, P16's process of labeling and retraining the model revealed a transformation in both their mental model and the model's learned patterns. Initially, P16 viewed the model's learned patterns for the label \textit{`sad'} as ``generic'', relying on rules such as `\textit{\$PERSON}' (matching all entities under person) and `\textit{PROPN|(mourn)}' (matching all sentences with proper nouns or synonyms of mourn). As the model was retrained with more targeted examples, P16 observed the emergence of more specific patterns, such as the rule \textit{(weep)++NOUN}, which matched phrases strictly involving synonyms of ``weep'' followed by a noun. 

While this was a common change in how the generated counterfactuals allowed the model to learn pattern rules with higher precision, some participants observed the model was giving correct labeling decisions but learning wrong patterns (i.e., write for the wrong reasons). For instance, P10 noticed that in the early stages of training, the model learned the pattern \textit{(sister)}\footnote{This pattern rule matches sentences that include any synonyms of the word \textit{sister}} for the label \textit{`fearful.'} Recognizing this as overfitting to the available labeled examples, P10 adjusted their strategy to focus on labeling counterfactuals. Specifically, they labeled examples where the pattern \textit{(sister)} occurred but were labeled with concepts other than fearful. This iterative approach eventually led the model to learn more appropriate patterns, such as \textit{(frighten)|(stand)} and \textit{(small)}, which aligned better with their intended concept of \textit{`fearful'}.}

The improvement in recall, although present, was more inconsistent across participants. In some cases, the integration of counterfactuals significantly enhanced the model's ability to identify relevant instances that were overlooked in the original dataset. For instance, P18's final model was able to learn the rule \textit{[sense] | (frightened)}\footnote{This pattern rule  matches all sentences with the literal word \textit{`sense'} or any synonyms of the word \textit{`frightened'}.} When trained with counterfactuals in addition to the original pattern rule \textit{(little) | (dread)}, it had learned with just the original examples for the label \textit{`fearful'} ultimately increasing the recall. Similarly, for the label \textit{`environment'} we observe the model's generated pattern rule change from \textit{(great)+(place)} into a combination of \textit{(atmosphere)|(area)} and \textit{(cozy)|[dining]} for P6. These cases illustrate how counterfactuals can enhance the flexibility and precision of pattern learning, leading to improvements in recall. However, the degree of improvement varied across different datasets and participants. Notably, the model performance for P18 actually decreased in recall, suggesting that the counterfactuals provided may have been less impactful, possibly because P18 had already achieved high performance without them. Further research is needed to explore these dynamics and understand the variable impacts of counterfactuals on model performance. 


Overall, the incorporation of counterfactuals has generally improved the models' F1 scores, driven largely by the improvements in precision. This suggests that counterfactuals have effectively improved performance without necessitating a significant trade-off between precision and recall. However, the less consistent increase in recall underscores the necessity for further research into how counterfactuals can be designed or selected more effectively to uniformly boost both aspects of model performance.

\subsubsection{\textbf{RQ3:} Co-adaptive learning with VT-based counterfactuals generation and Structural Alignment Theory-based counterfactual rendering}
\paragraph{Effects of alignable mapping in participant reading behavior}

Through the think-aloud sessions, we observe a noticeable shift in reading behavior when participants were annotating data points under C3, which came with highlighted alignable differences. In this condition, parts of the text were visually grayed out to denote redundant content from the original example, leading participants to generally skip these sections in their think-alouds. Instead, they concentrated on and vocalized sections of the text that remained bolded, representing new or added content in the generated counterfactual. This pattern of selective attention suggests that the visual cues provided by \toolname{} effectively guided participants to focus on more relevant information within the context of unchanged text when making their labeling decisions. \revision{It is worth noting that P3 and P8 mentioned feeling more comfortable with the more familiar visualization in C1 and C2 during their first impression of the conditions. When comparing their initial impression on the conditions, P8 said: \textit{``as I get familiar with this system [C3], I feel more skilled''} to use the highlighted and grayed phrases.}

\paragraph{Counterfactuals that followed Variation Theory enhanced model validation and retraining} 
Participants mentioned that counterfactuals adhering to Variation Theory significantly aided their understanding of the model's current learning state. Some participants used counterfactuals to validate the model’s understanding of specific labels. For example, P3 described their approach: \textit{``I think that is how I am using the counterfactuals to verify that it has actually learned what is the key part that makes this a product? And so if it just gives me [counterfactuals] [that are] probably still product, [then] maybe it has not learned anything about products. It has just learned things about other things''} (i.e., model overfitting). Alternatively, P5 observed that, if the generated counterfactuals belong to both the original and target labels,  the learned pattern rule might not be accurate enough (i.e., underfitting), remarking, \textit{``[the model] seems more confident''} when the generated counterfactuals belong exclusively to the target labels.

\paragraph{Highlighting alignable differences allowed participants to compare and contrast data points during annotation} Participants found that annotating counterfactuals---that were rendered with alignable differences highlighted---helpful in their decision-making. Specifically, two patterns of decision-making emerged. First, some participants~(P3, P4, P13) used the counterfactuals to revisit their interpretations of the original example and provide additional labels. For example, P4 adds a \textit{``product''} label to a sentence they previously labeled as \textit{``service''} after labeling the counterfactuals. \revision{In the followup interview, they reasoned that establishments that sell service as their products like a doctors office need to have the \textit{`product'} label as well.} Similarly, P3 stated, \textit{``I think by changing different parts of it [the original sentence], it highlighted a part of the sentence that I was not previously focused on, and so that did help me sort of reframe from what I initially had labeled it.''} By the end of the session, P3 reflected on how their initial interpretation of the label \textit{`service'} had changed: \textit{``I guess, going back to service, I did redefine, like we could also be talking about like the staff there, so like, including that in the labels.''} We also see participants similarly updating their understanding and interpretations of labels after seeing the counterfactuals: \textit{``it is more like the loose [definition of] happy we are talking about not just the word happy, but it sounds like it is just not like a vague mapping to the emotion''}~(P16).


Another decision-making behavior we saw consistently, perhaps an inverse of the previous one, was the participants' reliance on their original labeling to decide the counterfactuals' labels. This meant that the participants used the original example as an anchor for their consecutive interpretations. P18 used the original learned pattern rule for the label \textit{`fearful'} to provide an additional label for a generated counterfactual example. Specifically, after seeing that the generated counterfactual with the suggested label \textit{`sad'} also matched the pattern rule for the label \textit{`fearful'}, they annotated the sentence with both labels and stated, \textit{``I did not have that context [in the previous conditions] like I set up for myself, because reading this, I have like a story scenario in my head. [If] I did not have a scenario set up for myself I [would] probably just label it as only sad right away''}. 
The context provided in C3 allowed participants to infer additional labels to data points; this rarely happened in C1 (Non VT counterfactuals) where participants thought the generated examples were \textit{`independent'}~({P1}) from the original example and felt that they were labeling \textit{`completely new'}~({P18}) data points. In C2 (VT counterfactuals without highlighted alignable differences), despite following a similar approach to C3, participants struggled to identify differences between counterfactual examples, suggesting the effectiveness of highlighting structural consistency and differences in aiding comparative analysis.

  
  

  

%% file: 04-model_preformance_tab.tex
\begin{table*}[tbh!]
\centering
\begin{tabular}{@{}lrrr|rrr@{}}
\toprule
\multirow{2}{*}{PID} & \multicolumn{3}{c|}{\textbf{Without Counterexamples}} & \multicolumn{3}{c}{\textbf{With Counterexamples}} \\ \cmidrule(l){2-7} 
 & \multicolumn{1}{l}{F1-score} & \multicolumn{1}{l}{Precision} & \multicolumn{1}{l|}{Recall} & \multicolumn{1}{l}{F1-score} & \multicolumn{1}{l}{Precision} & \multicolumn{1}{l}{Recall} \\ \midrule
P1  & 0.43 &  0.52 & 0.37 & \textbf{0.63} & \textbf{0.85} & \textbf{0.50}\\
P2 & 0.47 & 0.63 & 0.38 & \textbf{0.72} & \textbf{0.90} & \textbf{0.67} \\
P3 & 0.64 & 0.60 & 0.70 & \textbf{0.80} & \textbf{0.82} & \textbf{0.83} \\
P4 & 0.46 & 0.65 & 0.37 & \textbf{0.85} & \textbf{0.93} & \textbf{0.80} \\
P5 & 0.45 & 0.45 & 0.45 & \textbf{0.78} & \textbf{0.90} & \textbf{0.69} \\
P6 & 0.61 & 0.75 & 0.55 & \textbf{0.84} & \textbf{0.98} & \textbf{0.75} \\
P7 & 0.40 & 0.46 & 0.35 & \textbf{0.66} & \textbf{0.91} & \textbf{0.54} \\
P8 & 0.51 & 0.75 & 0.42 & \textbf{0.55} & \textbf{0.70} & \textbf{0.46} \\
P9 & 0.64 & 0.72 & 0.61 & \textbf{0.77} & \textbf{0.91} & \textbf{0.69} \\
P10 & 0.45 & 0.45 & 0.46 & \textbf{0.85} & \textbf{0.92} & \textbf{0.80} \\
P11 & 0.42 & 0.47 & 0.39 & \textbf{0.60} & \textbf{0.75} & \textbf{0.52} \\
P12 & 0.38 & 0.48 & 0.31 & \textbf{0.63} & \textbf{0.72} & \textbf{0.56} \\
P13 & 0.57 & 0.68 & 0.52 & \textbf{0.75} & \textbf{0.89} & \textbf{0.69} \\
P14 & 0.41 & 0.60 & 0.32 & \textbf{0.73} & \textbf{0.92} & \textbf{0.60} \\
P15 & 0.44 & 0.62 & 0.36 & \textbf{0.74} & \textbf{0.91} & \textbf{0.66} \\
P16 & 0.54 & 0.67 & 0.46 & \textbf{0.67} & \textbf{0.90} & \textbf{0.54} \\
P17 & \textbf{0.68} & 0.86 & 0.59 & 0.67 & 0.86 & 0.59 \\
P18 & \textbf{0.92} & \textbf{0.96} & \textbf{0.90} & 0.78 & 0.94 & 0.67 \\ \bottomrule
\end{tabular}
\caption{The performance of the model with and without labeled counterfactuals for each participant}
\label{tab:mode-results}
\end{table*}

%% file: 05-discussion.tex
\section{Discussion and Design Implications}

\begin{figure}
    \centering
    \includegraphics[width=0.9\linewidth]{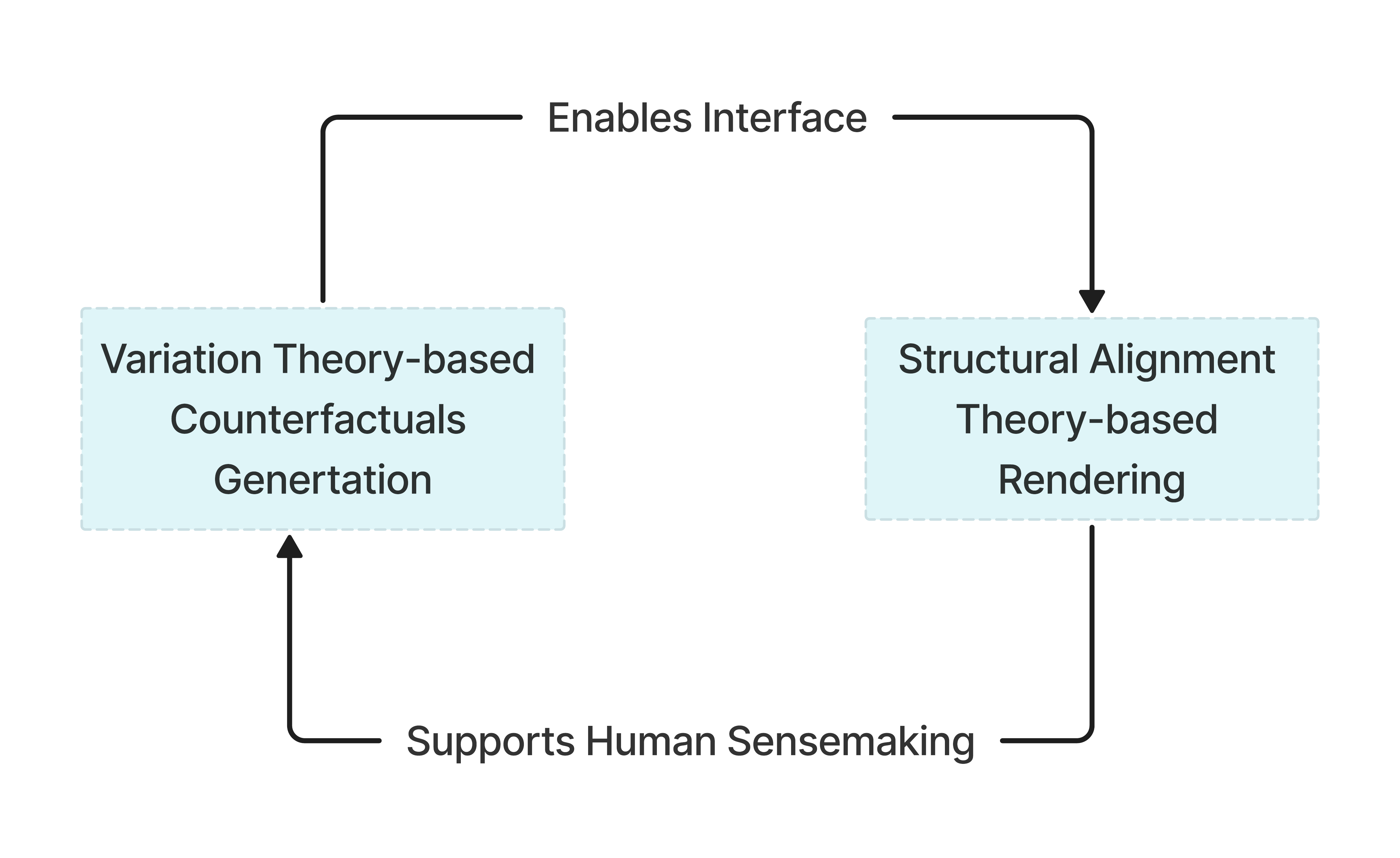}
    \Description{A diagram illustrating the relationship between Variation Theory-based Counterfactuals Generation and Structural Alignment Theory-based Rendering. The generation of counterfactuals based on Variation Theory enables the interface, which, in turn, uses Structural Alignment Theory for rendering. This interaction supports human sensemaking by facilitating a better understanding of data through alignable and non-alignable differences in the interface. The cycle shows how these two theories complement each other to enhance user interaction with the system.}
    \caption{Our study finds a bilateral relationship between Variation Theory and Structural Alignment Theory. The Variation Theory-based counterfactual generation method enabled the rendering of structurally alignable differences. In turn, the rendering supported the users' sensemaking of the variation in the generated counterfactuals}
    \label{fig:discussion-figure}
\end{figure}

The results from our user study suggest that both the participants and the model benefited from the Variation Theory (VT)-based counterfactuals and Structural Alignment Theory (SAT)-based rendering. Participants were able to efficiently focus on key differences between the original and counterfactual examples, which facilitated more efficient annotations. The participants also found \toolname{} to be useful in helping them refine and evolve their label definitions while giving them insight into the behavior of the model. Although the benefits of only Variation Theory-based counterfactuals (without SAT-based rendering) were not immediately evident in participants' experience, they were critical in enabling SAT-based rendering, which was found to be effective~(Fig.~\ref{fig:discussion-figure}). Below we further discuss our findings and design implications.

\subsection{A Symbiotic Relationship Between Variation Theory and Structural Alignment Theory}

The results of our study indicate that participants spent significantly less time annotating batches of counterfactuals when they were rendered according to SAT compared to other conditions i.e., supporting the participants' selective focus on the varying phrases, rather than phrases that stay consistent. Notably, we found no statistically significant differences in annotation time between C2~(VT-based counterfactual without SAT) and the baseline condition in C1. Although the counterfactuals in both C2 and C3 were based on Variation Theory, the participants' annotation efficiency was particularly enhanced by the SAT-based rendering method introduced in C3. 
\revision{This finding is consistent with previous work that supports users' sense-making of text, e.g., by modulating text saliency. Specifically, \citet{gu2024ai} and \citet{gero2024supporting} both found improved reading efficiency and comprehension with saliency-modulating text renderings. 
While SAT-based rendering supported human sense-making in both \citet{gero2024supporting} and \toolname{}, we also show that the combination of VT and SAT support the model's learning. }

To support a user's cognitive process for comparison~\cite{wisniewski1999makes}, the version of \toolname{} (C3) that followed a combination of Variation Theory and Structural Alignment Theory was consistently more effective. We argue that these two theories form a symbiotic relationship~(Fig.~\ref{fig:discussion-figure}). Variation Theory provides the conceptual basis for generating structurally consistent differences, while Structural Alignment Theory (SAT) enhances the user's ability in recognizing and processing these differences. This symbiotic relationship stems from the fact that Structural Alignment Theory (SAT) enhances the salience of differences, while the way we used Variation Theory (VT) to generate contradicting examples across the boundaries of labels ensures that these differences are conceptually informative. By helping users see alignable  differences, SAT-based rendering helps users focus on key variations that are essential to changing the data item's label, making it easier to interpret the effects of changes and their significance. Thus, the integration of both theories enables users to efficiently process and compare variations, leading to more informed decisions and a clearer understanding of the model's behavior.







\subsection{Human Cognition and Learning Theories in the Interactive ML Pipeline }

In its design, \toolname{} controls both how counterfactual data is generated and how they are rendered to the user. By incorporating theories such as Structural Alignment Theory and Variation Theory, it aims to support the learning of both the human and the model. These theories have proven insightful for understanding how humans grasp and compare concepts, shaping the development of human-AI collaboration systems for sensemaking~\cite{gero2024supporting}, hypothesis testing~\cite{arawjo2024chainforge}, as well as model training~\cite{geebregiabher2024leveraging}. From a human-in-the-loop machine learning perspective, \toolname{} addresses two seemingly contradictory objectives: (1) generating labeled data that diversifies the training dataset to aid the model's learning, and (2) maintaining structural consistency across the batches of data presented to users to support their cognitive processes. \toolname{} achieves this balance by enforcing a common structure through the model's learned pattern rules~\cite{gebreegziabher2023patat}. By visualizing these consistent pattern rules, users may be better understanding the behavior of the model through inference projection~\cite{gentner2010bootstrapping}. This can not only boosts the model's performance but also enable participants to validate or correct the model during the interactive training process.


Although visual cues for alignable differences in \toolname{} were helpful in supporting the participants' reasoning, \citet{estes2004importance} argue that while alignable differences can be more straightforward and easier for comparison, non-alignable differences can also provide key information that might otherwise remain overlooked. These differences necessitate a more abstract form of comparison, prompting users to think beyond simple relational structures and consider broader conceptual frameworks. For example, when comparing planes and cars, their alignable differences can be that they both have engines, but the engines are different in size. While this gives us some insight into their definitions, comparing a plane's wings and a car's wheels, which are not structurally alignable but conceptually and analogically alignable, gives additional insight to categorizing one for land and the other for air. Future research should explore how non-alignable differences in AI explanations affect user decision-making and understanding. Such studies could determine whether these non-alignable comparisons enhance user performance and elicit deeper insights in human-AI collaborative systems.




\section{Limitations and Future Work}
The current design of the user study, which allowed participants to interact with each condition for 25 minutes, yielded valuable insights into their immediate reactions and interactions. However, a longer study duration would provide a deeper understanding of how users engage with the system throughout various stages of the model's learning. Extending the study period would enable observations on how users' strategies evolve as the model improves, potentially uncovering significant insights about the long-term dynamics of human-AI collaboration, especially in relation to trust building and mental model refinement. Our study also only used example data from two domains, providing limited evidence to the approach's ecological validity in other data domains. In future work, we aim to investigate longer-term interactions of users in diverse application domains through deployment studies to uncover dynamic patterns of collaboration.

We also tested our system exclusively with the training of a neuro-symbolic model. While this serves as a compelling use case, it would be beneficial to explore the effectiveness of our approach with different types of models, such as purely statistical machine learning models, deep learning architectures, or hybrid systems, to better understand the generalizability of our approach across different paradigms. Different models might offer varying challenges and affordances in terms of explainability, interaction transparency, and feedback responsiveness. Exploring these dimensions would provide more insight into our proposed approach.

\section{Conclusion}
This paper introduced \toolname, an interactive machine learning tool informed by two theories of human concept learning and cognition. Based on the Variation Theory of human learning, it generates synthetic counterfactual data that are syntactically and semantically similar to already-annotated data but predicted by pre-trained large language models to have different labels. Following Structural Alignment Theory, it renders the generated counterfactuals aligned in batches with differences and similarities highlighted to support the user's cognitive process of interpreting and understanding data. A lab study with 18 participants demonstrated the usability of \toolname{} and its effectiveness in improving annotation efficiency, enhance the model’s learning, and facilitating co-adaptive learning in which users gain insight into the state of the model and reflect on their own understanding. \toolname{} exemplified the application of human cognition and concept learning theories in the interactive machine learning pipeline to support the negotiation of conceptual boundaries for bi-directional human-AI alignment.

%% file: 06-appendix.tex
\appendix
\section{Appendix}
\label{sec:appendix}

\subsection{User Study Participant Data}
\label{sec:study_participant_data}

\begin{table}[h]
\centering
\begin{tabular}{lllll}
\toprule
PID & Level of Education & Age &  Gender & ML Experience  \\ 
\midrule
P1  & Bachelor's &  18-24 &  Male & Beginner     \\ 
P2  & Bachelor's &  18-24 &  Male & Expert       \\ 
P3  & Bachelor's &  18-24 &  Male & Expert       \\ 
P4  & PhD        &  25-34 &  Female & Intermediate \\ 
P5  & Bachelor's &  25-34 &  Male & Intermediate \\ 
P6  & Master's   &  25-34 &  Male & Expert       \\ 
P7  & Master's   &  25-34 &  Male & Intermediate \\ 
P8  & Master's   &  25-34 &  Male & Intermediate \\ 
P9  & Master's   &  25-34 &  Male & Intermediate \\ 
P10 & Master's   &  25-34 &  Male & Intermediate \\ 
P11 & Master's   &  18-24 &  Male & Intermediate \\ 
P12 & Master's   &  25-34 &  Female & Beginner \\ 
P13 & Bachelor's   &  18-24 &  Female & Intermediate \\ 
P14 & Master's   &  25-34 &  Female & None \\ 
P15 & Bachelor's   &  18-24 &  Female & Intermediate \\ 
P16 & Master's  & 18-24  & Male   & Beginner \\ 
P17 & Master's   & 25-34 & Female  & Intermediate \\ 
P18 & Bachelor's   & 18-24  & Female  &  Beginner \\ 
\bottomrule
\end{tabular}
\caption{Participant demographic data}
\label{tab:participant-data}

\end{table}

\subsection{User Study Participant Task Details}
\label{sec:study_task_details}

\begin{table}[H]
\begin{tabular}{@{}lll@{}}
\toprule
Participant ID & Data & Condition Order \\ \midrule
P1 & Yelp & C1 - C3 - C2 \\
P2 & Yelp & C2 - C3 - C1 \\
P3 & Yelp & C3 - C1 - C2 \\
P4 & Yelp & C1 - C2 - C3 \\
P5 & Yelp & C1 - C3 - C2 \\
P6 & Yelp & C2 - C1 - C3 \\
P7 & Yelp & C3 - C2 - C1 \\
P8 & Yelp & C1 - C3 - C2 \\
P9 & Yelp & C3 - C1 - C2 \\
P10 & Emotions & C1 - C3- C2 \\
P11 & Emotions & C3 - C1 - C2 \\
P12 & Emotions & C2 - C1 - C3 \\
P13 & Emotions & C2 - C3 - C1 \\
P14 & Emotions & C3 - C1 - C2 \\
P15 & Emotions & C3 - C2 - C1 \\
P16 & Emotions & C1 - C2 - C3 \\
P17 & Emotions & C1 - C3 - C2 \\
P18 & Emotions & C2 - C1 - C3 \\ \bottomrule
\end{tabular}
\caption{User study participant's assigned data and condition order}
\label{tab:task-details}
\end{table}


\subsection{Neuro-symbolic Pattern Rules}
\label{app:patat-rules}
\revision{The neuro-symbolic model adopts an iterative learning approach to delineate the boundaries of concepts represented by data points and their corresponding ground truth labels. Although the current simulation study employs ground truth labels, these will eventually be replaced with human annotations in future interactive systems. Following a random selection of a subset of annotations, the interactive program synthesis method from PaTAT~\cite{gebreegziabher2023patat} is applied to derive domain-specific pattern rules that align with the annotated examples. These rules capture the lexical, syntactic, and semantic similarities present among data sharing the same label. The pattern language is composed of the following components:}

\revision{\begin{itemize}
    \item Part-of-speech (POS) tags: \texttt{VERB}, \texttt{PROPN}, \texttt{NOUN}, \texttt{ADJ}, \texttt{ADV}, \texttt{AUX}, \texttt{PRON}, \texttt{NUM}
    \item Word stemming: \texttt{[WORD]} (e.g., \texttt{[have]} will match all variants of have, such as \textit{had}, \textit{has}, and \textit{having})
    \item Soft match: \texttt{(word)} (e.g., \texttt{(pricey)} will match synonyms such as \textit{expensive} and \textit{costly}, etc.)
    \item Entity type: \texttt{\$ENT-TYPE} (e.g., \texttt{\$LOCATION} will match phrases of location type, such as \textit{Houston, TX} and \textit{California}; \texttt{\$DATE} will match dates; \texttt{\$ORG} will match names of organizations)
    \item Wildcard: \texttt{*} (will match any sequence of words)
\end{itemize}}

\subsection{Candidate Phrase Generation Prompt}
\label{sec:candidate-phrases}
\begin{lstlisting}
The assistant will create a list of candidate phrases that match the given symbolic domain specific pattern. The domain specific pattern definition is given below. The domain specific pattern symbols includes the following patterns:
Part-of-speech (POS) tags are capital: VERB, PROPN, NOUN, ADJ, ADV, AUX, PRON, NUM
Word stemming are surrounded in [] and should have an exact match: [WORD] (e.g., [have] will match all variants of have)
Soft match are surrounded by () and will match words with their synonyms. The list of synonms for each soft match in a pattern are given in the user instruction: (word) (word will only be matched with a limited set of similar words provided in this instruction)
Entity type start with $ sign: $ENT-TYPE (e.g., $LOCATION will match phrases of location type, such as Houston; $DATE will match dates)
Wildcard is the * symbol and can match anything: * (will match any sequence of words)
The patterns can be combined using an and operator (+) or an or operator (|). For example the pattern 'VERB + PROPN' will match any sentence that has a verb followed by a proper noun. The pattern VERB|PROPN will match anything that is a verb or a proper noun.
Soft matches can only be replaced with a list of available words.
For the following text and pattern, generate as many diverse example phrases that match the given pattern and can be part of the given target label. Separated your answer by a comma
# Example input:
# sentence: 'Too many other places to shop with better prices .'
# phrase to modify: 'prices .'
# pattern: '(price)+*'
# current label: price
# softmatch: [price:[purchase, pricey, cheap, cost, pricing]]
# target label: service
# Example output:
# [purchase options, pricey service, cheap help, pricing plans, cost breakdown]}
\end{lstlisting}

\subsection{Counterfactual Generation Prompt}
\label{sec:counter-generation}
\begin{lstlisting}
Your task is to modify a given sentence so that it aligns with a target label instead of its original label, making only necessary changes. Follow these steps:
- Generate Target Phrases: Identify phrases relevant to the target label within the context of the original sentence.
- Modify the Sentence: Use one of the generated target phrases to adjust the original sentence, ensuring that:
- The modified sentence no longer fits the original label and does not reference or imply the original label in any way.
- The modified sentence is appropriate for the target label and logically coherent.
- The modified sentence should be natural and fluent, making sense as a standalone sentence.
- Changes made are necessary while keeping the original sentence structure as intact as possible. To preserve the quality of the new sentence you can remove or add necessary parts.
- The modification includes one of the provided candidate phrases, replacing the highlighted portion of the original sentence.
- Explanation (Optional): If necessary, provide a brief explanation of why the modified sentence fits the target label.
Explanation of Terms:
* phrase about target label: generated phrases relevant to the target label that help guide the sentence modification.
* phrases to include: this includes one phrase from 'phrase about target label' and another phrase from the user input 'candidate phrases'. these two phrases will be incorporated into the modified sentence.
* modified sentence: The final sentence after modification to align with the target label. It should be natural, logical, and coherent.
* reason: A brief explanation of why the modified sentence fits the target label.
* label: The final label assigned to the modified sentence, reflecting its changes.
* Ensure the final sentence is correctly labeled according to the target label, with no references to the original label, and with minimal deviation from the original content.
# Example input:
# Original sentence: 'The wings were delicious.'
# Original label: product,
# Target label: price,
# Candidate phrases: ['yummy', 'tasty', 'flavour', 'deliciousness', 'taste', 'delicious']
# phrase about target label: ['cheap', 'expensive', 'pricey']
# phrase to include: ['taste', 'cheap']
# Example output:
# modified sentence: 'The wings were cheap for the taste.'
# reason: 'The sentence shifts focus to the cost of the wings, making it fit the target label price.'
# label: price
\end{lstlisting}


\subsection{NASA-TLX Results}
\label{sec:nasa-tlx}

\begin{figure*}[b]
    \centering
    \includegraphics[width=\linewidth]{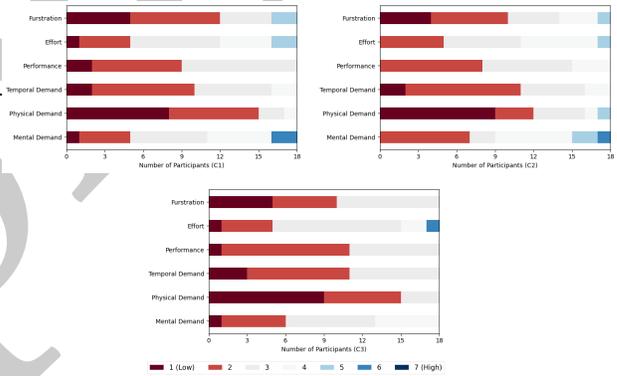}
    \Description{The figure presents stacked bar charts comparing participant ratings for six categories: Frustration, Effort, Performance, Temporal Demand, Physical Demand, and Mental Demand across three conditions}
    \caption{NASA-TLX results from the user study}
    \label{fig:nasa-tlx}
\end{figure*}

%% file: main.bbl

\begin{thebibliography}{69}


\ifx \showCODEN    \undefined \def \showCODEN     #1{\unskip}     \fi
\ifx \showDOI      \undefined \def \showDOI       #1{#1}\fi
\ifx \showISBNx    \undefined \def \showISBNx     #1{\unskip}     \fi
\ifx \showISBNxiii \undefined \def \showISBNxiii  #1{\unskip}     \fi
\ifx \showISSN     \undefined \def \showISSN      #1{\unskip}     \fi
\ifx \showLCCN     \undefined \def \showLCCN      #1{\unskip}     \fi
\ifx \shownote     \undefined \def \shownote      #1{#1}          \fi
\ifx \showarticletitle \undefined \def \showarticletitle #1{#1}   \fi
\ifx \showURL      \undefined \def \showURL       {\relax}        \fi
\providecommand\bibfield[2]{#2}
\providecommand\bibinfo[2]{#2}
\providecommand\natexlab[1]{#1}
\providecommand\showeprint[2][]{arXiv:#2}

\bibitem[Alm(2008)]%
        {alm2008affect}
\bibfield{author}{\bibinfo{person}{Ebba Cecilia~Ovesdotter Alm}.} \bibinfo{year}{2008}\natexlab{}.
\newblock \bibinfo{booktitle}{\emph{Affect in* text and speech}}.
\newblock \bibinfo{publisher}{University of Illinois at Urbana-Champaign}.
\newblock


\bibitem[Arawjo et~al\mbox{.}(2024)]%
        {arawjo2024chainforge}
\bibfield{author}{\bibinfo{person}{Ian Arawjo}, \bibinfo{person}{Chelse Swoopes}, \bibinfo{person}{Priyan Vaithilingam}, \bibinfo{person}{Martin Wattenberg}, {and} \bibinfo{person}{Elena~L Glassman}.} \bibinfo{year}{2024}\natexlab{}.
\newblock \showarticletitle{ChainForge: A Visual Toolkit for Prompt Engineering and LLM Hypothesis Testing}. In \bibinfo{booktitle}{\emph{Proceedings of the CHI Conference on Human Factors in Computing Systems}}. \bibinfo{pages}{1--18}.
\newblock


\bibitem[Arora and Agarwal(2007)]%
        {arora2007active}
\bibfield{author}{\bibinfo{person}{Shilpa Arora} {and} \bibinfo{person}{Sachin Agarwal}.} \bibinfo{year}{2007}\natexlab{}.
\newblock \showarticletitle{Active learning for natural language processing}.
\newblock \bibinfo{journal}{\emph{Language Technologies Institute School of Computer Science Carnegie Mellon University}}  \bibinfo{volume}{2} (\bibinfo{year}{2007}).
\newblock


\bibitem[Aroyo et~al\mbox{.}(2022)]%
        {aroyo2022data}
\bibfield{author}{\bibinfo{person}{Lora Aroyo}, \bibinfo{person}{Matthew Lease}, \bibinfo{person}{Praveen Paritosh}, {and} \bibinfo{person}{Mike Schaekermann}.} \bibinfo{year}{2022}\natexlab{}.
\newblock \showarticletitle{Data excellence for AI: why should you care?}
\newblock \bibinfo{journal}{\emph{Interactions}} \bibinfo{volume}{29}, \bibinfo{number}{2} (\bibinfo{year}{2022}), \bibinfo{pages}{66--69}.
\newblock


\bibitem[Asghar(2016)]%
        {asghar2016yelp}
\bibfield{author}{\bibinfo{person}{Nabiha Asghar}.} \bibinfo{year}{2016}\natexlab{}.
\newblock \showarticletitle{Yelp dataset challenge: Review rating prediction}.
\newblock \bibinfo{journal}{\emph{arXiv preprint arXiv:1605.05362}} (\bibinfo{year}{2016}).
\newblock


\bibitem[Booth et~al\mbox{.}(2022)]%
        {booth2022revisiting}
\bibfield{author}{\bibinfo{person}{Serena Booth}, \bibinfo{person}{Sanjana Sharma}, \bibinfo{person}{Sarah Chung}, \bibinfo{person}{Julie Shah}, {and} \bibinfo{person}{Elena~L Glassman}.} \bibinfo{year}{2022}\natexlab{}.
\newblock \showarticletitle{Revisiting human-robot teaching and learning through the lens of human concept learning}. In \bibinfo{booktitle}{\emph{2022 17th ACM/IEEE International Conference on Human-Robot Interaction (HRI)}}. IEEE, \bibinfo{pages}{147--156}.
\newblock


\bibitem[Brooks et~al\mbox{.}(2015)]%
        {brooks2015featureinsight}
\bibfield{author}{\bibinfo{person}{Michael Brooks}, \bibinfo{person}{Saleema Amershi}, \bibinfo{person}{Bongshin Lee}, \bibinfo{person}{Steven~M Drucker}, \bibinfo{person}{Ashish Kapoor}, {and} \bibinfo{person}{Patrice Simard}.} \bibinfo{year}{2015}\natexlab{}.
\newblock \showarticletitle{FeatureInsight: Visual support for error-driven feature ideation in text classification}. In \bibinfo{booktitle}{\emph{2015 IEEE Conference on Visual Analytics Science and Technology (VAST)}}. IEEE, \bibinfo{pages}{105--112}.
\newblock


\bibitem[Bunzel et~al\mbox{.}(2024)]%
        {bunzel2024identifying}
\bibfield{author}{\bibinfo{person}{Niklas Bunzel}, \bibinfo{person}{Nicolas G{\"o}ller}, {and} \bibinfo{person}{Raphael~Antonius Frick}.} \bibinfo{year}{2024}\natexlab{}.
\newblock \showarticletitle{Identifying and Generating Edge Cases}. In \bibinfo{booktitle}{\emph{Proceedings of the 2nd ACM Workshop on Secure and Trustworthy Deep Learning Systems}}. \bibinfo{pages}{16--23}.
\newblock


\bibitem[Cabrera et~al\mbox{.}(2023)]%
        {cabrera2023did}
\bibfield{author}{\bibinfo{person}{{\'A}ngel~Alexander Cabrera}, \bibinfo{person}{Marco Tulio~Ribeiro}, \bibinfo{person}{Bongshin Lee}, \bibinfo{person}{Robert Deline}, \bibinfo{person}{Adam Perer}, {and} \bibinfo{person}{Steven~M Drucker}.} \bibinfo{year}{2023}\natexlab{}.
\newblock \showarticletitle{What did my AI learn? How data scientists make sense of model behavior}.
\newblock \bibinfo{journal}{\emph{ACM Transactions on Computer-Human Interaction}} \bibinfo{volume}{30}, \bibinfo{number}{1} (\bibinfo{year}{2023}), \bibinfo{pages}{1--27}.
\newblock


\bibitem[Chang et~al\mbox{.}(2016)]%
        {chang2016alloy}
\bibfield{author}{\bibinfo{person}{Joseph~Chee Chang}, \bibinfo{person}{Aniket Kittur}, {and} \bibinfo{person}{Nathan Hahn}.} \bibinfo{year}{2016}\natexlab{}.
\newblock \showarticletitle{Alloy: Clustering with crowds and computation}. In \bibinfo{booktitle}{\emph{Proceedings of the 2016 CHI Conference on Human Factors in Computing Systems}}. \bibinfo{pages}{3180--3191}.
\newblock


\bibitem[Chen et~al\mbox{.}(2022)]%
        {chen2022disco}
\bibfield{author}{\bibinfo{person}{Zeming Chen}, \bibinfo{person}{Qiyue Gao}, \bibinfo{person}{Antoine Bosselut}, \bibinfo{person}{Ashish Sabharwal}, {and} \bibinfo{person}{Kyle Richardson}.} \bibinfo{year}{2022}\natexlab{}.
\newblock \showarticletitle{DISCO: Distilling counterfactuals with large language models}.
\newblock \bibinfo{journal}{\emph{arXiv preprint arXiv:2212.10534}} (\bibinfo{year}{2022}).
\newblock


\bibitem[Cruz and Sunna(2008)]%
        {cruz2008structural}
\bibfield{author}{\bibinfo{person}{Isabel~F Cruz} {and} \bibinfo{person}{William Sunna}.} \bibinfo{year}{2008}\natexlab{}.
\newblock \showarticletitle{Structural alignment methods with applications to geospatial ontologies}.
\newblock \bibinfo{journal}{\emph{Transactions in GIS}} \bibinfo{volume}{12}, \bibinfo{number}{6} (\bibinfo{year}{2008}), \bibinfo{pages}{683--711}.
\newblock


\bibitem[Culotta and McCallum(2005)]%
        {culotta2005reducing}
\bibfield{author}{\bibinfo{person}{Aron Culotta} {and} \bibinfo{person}{Andrew McCallum}.} \bibinfo{year}{2005}\natexlab{}.
\newblock \showarticletitle{Reducing labeling effort for structured prediction tasks}. In \bibinfo{booktitle}{\emph{AAAI}}, Vol.~\bibinfo{volume}{5}. \bibinfo{pages}{746--751}.
\newblock


\bibitem[Dixit et~al\mbox{.}(2022)]%
        {dixit2022core}
\bibfield{author}{\bibinfo{person}{Tanay Dixit}, \bibinfo{person}{Bhargavi Paranjape}, \bibinfo{person}{Hannaneh Hajishirzi}, {and} \bibinfo{person}{Luke Zettlemoyer}.} \bibinfo{year}{2022}\natexlab{}.
\newblock \showarticletitle{CORE: A retrieve-then-edit framework for counterfactual data generation}.
\newblock \bibinfo{journal}{\emph{arXiv preprint arXiv:2210.04873}} (\bibinfo{year}{2022}).
\newblock


\bibitem[Dodge et~al\mbox{.}(2020)]%
        {dodge2020fine}
\bibfield{author}{\bibinfo{person}{Jesse Dodge}, \bibinfo{person}{Gabriel Ilharco}, \bibinfo{person}{Roy Schwartz}, \bibinfo{person}{Ali Farhadi}, \bibinfo{person}{Hannaneh Hajishirzi}, {and} \bibinfo{person}{Noah Smith}.} \bibinfo{year}{2020}\natexlab{}.
\newblock \showarticletitle{Fine-tuning pretrained language models: Weight initializations, data orders, and early stopping}.
\newblock \bibinfo{journal}{\emph{arXiv preprint arXiv:2002.06305}} (\bibinfo{year}{2020}).
\newblock


\bibitem[Dudley and Kristensson(2018)]%
        {dudley2018review}
\bibfield{author}{\bibinfo{person}{John~J Dudley} {and} \bibinfo{person}{Per~Ola Kristensson}.} \bibinfo{year}{2018}\natexlab{}.
\newblock \showarticletitle{A review of user interface design for interactive machine learning}.
\newblock \bibinfo{journal}{\emph{ACM Transactions on Interactive Intelligent Systems (TiiS)}} \bibinfo{volume}{8}, \bibinfo{number}{2} (\bibinfo{year}{2018}), \bibinfo{pages}{1--37}.
\newblock


\bibitem[Estes and Hasson(2004)]%
        {estes2004importance}
\bibfield{author}{\bibinfo{person}{Zachary Estes} {and} \bibinfo{person}{Uri Hasson}.} \bibinfo{year}{2004}\natexlab{}.
\newblock \showarticletitle{The importance of being nonalignable: a critical test of the structural alignment theory of similarity.}
\newblock \bibinfo{journal}{\emph{Journal of Experimental Psychology: Learning, Memory, and Cognition}} \bibinfo{volume}{30}, \bibinfo{number}{5} (\bibinfo{year}{2004}), \bibinfo{pages}{1082}.
\newblock


\bibitem[Fails and Olsen~Jr(2003)]%
        {fails2003interactive}
\bibfield{author}{\bibinfo{person}{Jerry~Alan Fails} {and} \bibinfo{person}{Dan~R Olsen~Jr}.} \bibinfo{year}{2003}\natexlab{}.
\newblock \showarticletitle{Interactive machine learning}. In \bibinfo{booktitle}{\emph{Proceedings of the 8th international conference on Intelligent user interfaces}}. \bibinfo{pages}{39--45}.
\newblock


\bibitem[Feder et~al\mbox{.}(2021)]%
        {feder2021causalm}
\bibfield{author}{\bibinfo{person}{Amir Feder}, \bibinfo{person}{Nadav Oved}, \bibinfo{person}{Uri Shalit}, {and} \bibinfo{person}{Roi Reichart}.} \bibinfo{year}{2021}\natexlab{}.
\newblock \showarticletitle{Causalm: Causal model explanation through counterfactual language models}.
\newblock \bibinfo{journal}{\emph{Computational Linguistics}} \bibinfo{volume}{47}, \bibinfo{number}{2} (\bibinfo{year}{2021}), \bibinfo{pages}{333--386}.
\newblock


\bibitem[Felder and Brent(2009)]%
        {felder2009active}
\bibfield{author}{\bibinfo{person}{Richard~M Felder} {and} \bibinfo{person}{Rebecca Brent}.} \bibinfo{year}{2009}\natexlab{}.
\newblock \showarticletitle{Active learning: An introduction}.
\newblock \bibinfo{journal}{\emph{ASQ higher education brief}} \bibinfo{volume}{2}, \bibinfo{number}{4} (\bibinfo{year}{2009}), \bibinfo{pages}{1--5}.
\newblock


\bibitem[Field(2005)]%
        {field2005kendall}
\bibfield{author}{\bibinfo{person}{Andy~P Field}.} \bibinfo{year}{2005}\natexlab{}.
\newblock \showarticletitle{Kendall's coefficient of concordance}.
\newblock \bibinfo{journal}{\emph{Encyclopedia of statistics in behavioral science}}  \bibinfo{volume}{2} (\bibinfo{year}{2005}), \bibinfo{pages}{1010--11}.
\newblock


\bibitem[Fleiss et~al\mbox{.}(1969)]%
        {fleiss1969large}
\bibfield{author}{\bibinfo{person}{Joseph~L Fleiss}, \bibinfo{person}{Jacob Cohen}, {and} \bibinfo{person}{Brian~S Everitt}.} \bibinfo{year}{1969}\natexlab{}.
\newblock \showarticletitle{Large sample standard errors of kappa and weighted kappa.}
\newblock \bibinfo{journal}{\emph{Psychological bulletin}} \bibinfo{volume}{72}, \bibinfo{number}{5} (\bibinfo{year}{1969}), \bibinfo{pages}{323}.
\newblock


\bibitem[Friedman(1937)]%
        {friedman1937use}
\bibfield{author}{\bibinfo{person}{Milton Friedman}.} \bibinfo{year}{1937}\natexlab{}.
\newblock \showarticletitle{The use of ranks to avoid the assumption of normality implicit in the analysis of variance}.
\newblock \bibinfo{journal}{\emph{Journal of the american statistical association}} \bibinfo{volume}{32}, \bibinfo{number}{200} (\bibinfo{year}{1937}), \bibinfo{pages}{675--701}.
\newblock


\bibitem[Gebreegziabher et~al\mbox{.}(2024)]%
        {geebregiabher2024leveraging}
\bibfield{author}{\bibinfo{person}{Simret~Araya Gebreegziabher}, \bibinfo{person}{Kuangshi Ai}, \bibinfo{person}{Zheng Zhang}, \bibinfo{person}{Elena~L. Glassman}, {and} \bibinfo{person}{Toby Jia-Jun Li}.} \bibinfo{year}{2024}\natexlab{}.
\newblock \showarticletitle{Leveraging Variation Theory in Counterfactual Data Augmentation for Optimized Active Learning}.
\newblock \bibinfo{journal}{\emph{arXiv preprint arXiv:2408.03819}} (\bibinfo{year}{2024}).
\newblock


\bibitem[Gebreegziabher et~al\mbox{.}(2023)]%
        {gebreegziabher2023patat}
\bibfield{author}{\bibinfo{person}{Simret~Araya Gebreegziabher}, \bibinfo{person}{Zheng Zhang}, \bibinfo{person}{Xiaohang Tang}, \bibinfo{person}{Yihao Meng}, \bibinfo{person}{Elena~L Glassman}, {and} \bibinfo{person}{Toby Jia-Jun Li}.} \bibinfo{year}{2023}\natexlab{}.
\newblock \showarticletitle{Patat: Human-ai collaborative qualitative coding with explainable interactive rule synthesis}. In \bibinfo{booktitle}{\emph{Proceedings of the 2023 CHI Conference on Human Factors in Computing Systems}}. \bibinfo{pages}{1--19}.
\newblock


\bibitem[Gentner(2010)]%
        {gentner2010bootstrapping}
\bibfield{author}{\bibinfo{person}{Dedre Gentner}.} \bibinfo{year}{2010}\natexlab{}.
\newblock \showarticletitle{Bootstrapping the mind: Analogical processes and symbol systems}.
\newblock \bibinfo{journal}{\emph{Cognitive science}} \bibinfo{volume}{34}, \bibinfo{number}{5} (\bibinfo{year}{2010}), \bibinfo{pages}{752--775}.
\newblock


\bibitem[Gentner and Gunn(2001)]%
        {gentner2001structural}
\bibfield{author}{\bibinfo{person}{Dedre Gentner} {and} \bibinfo{person}{Virginia Gunn}.} \bibinfo{year}{2001}\natexlab{}.
\newblock \showarticletitle{Structural alignment facilitates the noticing of differences}.
\newblock \bibinfo{journal}{\emph{Memory \& cognition}} \bibinfo{volume}{29}, \bibinfo{number}{4} (\bibinfo{year}{2001}), \bibinfo{pages}{565--577}.
\newblock


\bibitem[Gentner and Markman(1997)]%
        {gentner1997structure}
\bibfield{author}{\bibinfo{person}{Dedre Gentner} {and} \bibinfo{person}{Arthur~B Markman}.} \bibinfo{year}{1997}\natexlab{}.
\newblock \showarticletitle{Structure mapping in analogy and similarity}.
\newblock \bibinfo{journal}{\emph{American psychologist}} \bibinfo{volume}{52}, \bibinfo{number}{1} (\bibinfo{year}{1997}), \bibinfo{pages}{45}.
\newblock


\bibitem[Gero et~al\mbox{.}(2024)]%
        {gero2024supporting}
\bibfield{author}{\bibinfo{person}{Katy~Ilonka Gero}, \bibinfo{person}{Chelse Swoopes}, \bibinfo{person}{Ziwei Gu}, \bibinfo{person}{Jonathan~K Kummerfeld}, {and} \bibinfo{person}{Elena~L Glassman}.} \bibinfo{year}{2024}\natexlab{}.
\newblock \showarticletitle{Supporting Sensemaking of Large Language Model Outputs at Scale}. In \bibinfo{booktitle}{\emph{Proceedings of the CHI Conference on Human Factors in Computing Systems}}. \bibinfo{pages}{1--21}.
\newblock


\bibitem[Gillies et~al\mbox{.}(2016)]%
        {gillies2016human}
\bibfield{author}{\bibinfo{person}{Marco Gillies}, \bibinfo{person}{Rebecca Fiebrink}, \bibinfo{person}{Atau Tanaka}, \bibinfo{person}{J{\'e}r{\'e}mie Garcia}, \bibinfo{person}{Fr{\'e}d{\'e}ric Bevilacqua}, \bibinfo{person}{Alexis Heloir}, \bibinfo{person}{Fabrizio Nunnari}, \bibinfo{person}{Wendy Mackay}, \bibinfo{person}{Saleema Amershi}, \bibinfo{person}{Bongshin Lee}, {et~al\mbox{.}}} \bibinfo{year}{2016}\natexlab{}.
\newblock \showarticletitle{Human-centred machine learning}. In \bibinfo{booktitle}{\emph{Proceedings of the 2016 CHI conference extended abstracts on human factors in computing systems}}. \bibinfo{pages}{3558--3565}.
\newblock


\bibitem[Gomes et~al\mbox{.}(2024)]%
        {gomes2024finding}
\bibfield{author}{\bibinfo{person}{In{\^e}s Gomes}, \bibinfo{person}{Lu{\'\i}s~F Teixeira}, \bibinfo{person}{Jan~N Van~Rijn}, \bibinfo{person}{Carlos Soares}, \bibinfo{person}{Andr{\'e} Restivo}, \bibinfo{person}{Lu{\'\i}s Cunha}, {and} \bibinfo{person}{Mois{\'e}s Santos}.} \bibinfo{year}{2024}\natexlab{}.
\newblock \showarticletitle{Finding Patterns in Ambiguity: Interpretable Stress Testing in the Decision Boundary}. In \bibinfo{booktitle}{\emph{Proceedings of the IEEE/CVF Conference on Computer Vision and Pattern Recognition}}. \bibinfo{pages}{8316--8321}.
\newblock


\bibitem[Gu et~al\mbox{.}(2024)]%
        {gu2024ai}
\bibfield{author}{\bibinfo{person}{Ziwei Gu}, \bibinfo{person}{Ian Arawjo}, \bibinfo{person}{Kenneth Li}, \bibinfo{person}{Jonathan~K Kummerfeld}, {and} \bibinfo{person}{Elena~L Glassman}.} \bibinfo{year}{2024}\natexlab{}.
\newblock \showarticletitle{An AI-Resilient Text Rendering Technique for Reading and Skimming Documents}. In \bibinfo{booktitle}{\emph{Proceedings of the CHI Conference on Human Factors in Computing Systems}}. \bibinfo{pages}{1--22}.
\newblock


\bibitem[Habibelahian et~al\mbox{.}(2022)]%
        {habibelahian2022exploratory}
\bibfield{author}{\bibinfo{person}{Omeed Habibelahian}, \bibinfo{person}{Rajesh Shrestha}, \bibinfo{person}{Arash Termehchy}, {and} \bibinfo{person}{Paolo Papotti}.} \bibinfo{year}{2022}\natexlab{}.
\newblock \showarticletitle{Exploratory training: when trainers learn}. In \bibinfo{booktitle}{\emph{Proceedings of the Workshop on Human-In-the-Loop Data Analytics}}. \bibinfo{pages}{1--5}.
\newblock


\bibitem[Hart(2006)]%
        {hart2006nasa}
\bibfield{author}{\bibinfo{person}{Sandra~G Hart}.} \bibinfo{year}{2006}\natexlab{}.
\newblock \showarticletitle{NASA-task load index (NASA-TLX); 20 years later}. In \bibinfo{booktitle}{\emph{Proceedings of the human factors and ergonomics society annual meeting}}, Vol.~\bibinfo{volume}{50}. Sage publications Sage CA: Los Angeles, CA, \bibinfo{pages}{904--908}.
\newblock


\bibitem[Hassan and Alikhani(2023)]%
        {hassan2023d}
\bibfield{author}{\bibinfo{person}{Sabit Hassan} {and} \bibinfo{person}{Malihe Alikhani}.} \bibinfo{year}{2023}\natexlab{}.
\newblock \showarticletitle{D-CALM: A dynamic clustering-based active learning approach for mitigating bias}.
\newblock \bibinfo{journal}{\emph{arXiv preprint arXiv:2305.17013}} (\bibinfo{year}{2023}).
\newblock


\bibitem[Jacovi(2021)]%
        {Jacovi2021Formalizing}
\bibfield{author}{\bibinfo{person}{Marasović Jacovi}.} \bibinfo{year}{2021}\natexlab{}.
\newblock \showarticletitle{Formalizing Trust in Artificial Intelligence: Prerequisites, Causes and Goals of Human Trust in AI}.
\newblock \bibinfo{journal}{\emph{URL https://dl.acm.org/doi/10.1145/3442188.3445923}} (\bibinfo{year}{2021}).
\newblock


\bibitem[Kivij{\"a}rvi(2018)]%
        {kivijarvi2018advancing}
\bibfield{author}{\bibinfo{person}{Hannu Kivij{\"a}rvi}.} \bibinfo{year}{2018}\natexlab{}.
\newblock \showarticletitle{Advancing Organizational Alignment Decisions: Insights from the Structural Alignment Theory to the Business-IT Alignment Problem}.
\newblock \bibinfo{journal}{\emph{International Journal of IT/Business Alignment and Governance (IJITBAG)}} \bibinfo{volume}{9}, \bibinfo{number}{1} (\bibinfo{year}{2018}), \bibinfo{pages}{53--80}.
\newblock


\bibitem[Konzett-Firth(2020)]%
        {konzett2020co}
\bibfield{author}{\bibinfo{person}{Carmen Konzett-Firth}.} \bibinfo{year}{2020}\natexlab{}.
\newblock \showarticletitle{Co-adaptation processes in plenary teacher-student talk and the development of L2 interactional competence}.
\newblock \bibinfo{journal}{\emph{Classroom Discourse}} \bibinfo{volume}{11}, \bibinfo{number}{3} (\bibinfo{year}{2020}), \bibinfo{pages}{209--228}.
\newblock


\bibitem[Kruskal and Wallis(1952)]%
        {kruskal1952use}
\bibfield{author}{\bibinfo{person}{William~H Kruskal} {and} \bibinfo{person}{W~Allen Wallis}.} \bibinfo{year}{1952}\natexlab{}.
\newblock \showarticletitle{Use of ranks in one-criterion variance analysis}.
\newblock \bibinfo{journal}{\emph{Journal of the American statistical Association}} \bibinfo{volume}{47}, \bibinfo{number}{260} (\bibinfo{year}{1952}), \bibinfo{pages}{583--621}.
\newblock


\bibitem[Kulesza et~al\mbox{.}(2014)]%
        {kulesza2014structured}
\bibfield{author}{\bibinfo{person}{Todd Kulesza}, \bibinfo{person}{Saleema Amershi}, \bibinfo{person}{Rich Caruana}, \bibinfo{person}{Danyel Fisher}, {and} \bibinfo{person}{Denis Charles}.} \bibinfo{year}{2014}\natexlab{}.
\newblock \showarticletitle{Structured labeling for facilitating concept evolution in machine learning}. In \bibinfo{booktitle}{\emph{Proceedings of the SIGCHI Conference on Human Factors in Computing Systems}}. \bibinfo{pages}{3075--3084}.
\newblock


\bibitem[Lee et~al\mbox{.}(2012)]%
        {lee2012based}
\bibfield{author}{\bibinfo{person}{Chi-Chun Lee}, \bibinfo{person}{Athanasios Katsamanis}, \bibinfo{person}{Panayiotis~G Georgiou}, {and} \bibinfo{person}{Shrikanth~S Narayanan}.} \bibinfo{year}{2012}\natexlab{}.
\newblock \showarticletitle{Based on Isolated Saliency or Causal Integration? Toward a Better Understanding of Human Annotation Process using Multiple Instance Learning and Sequential Probability Ratio Test.}. In \bibinfo{booktitle}{\emph{INTERSPEECH}}. \bibinfo{pages}{619--622}.
\newblock


\bibitem[Lewis(1995)]%
        {lewis1995sequential}
\bibfield{author}{\bibinfo{person}{David~D Lewis}.} \bibinfo{year}{1995}\natexlab{}.
\newblock \showarticletitle{A sequential algorithm for training text classifiers: Corrigendum and additional data}. In \bibinfo{booktitle}{\emph{Acm Sigir Forum}}, Vol.~\bibinfo{volume}{29}. ACM New York, NY, USA, \bibinfo{pages}{13--19}.
\newblock


\bibitem[Margatina et~al\mbox{.}(2021)]%
        {margatina2021active}
\bibfield{author}{\bibinfo{person}{Katerina Margatina}, \bibinfo{person}{Giorgos Vernikos}, \bibinfo{person}{Lo{\"\i}c Barrault}, {and} \bibinfo{person}{Nikolaos Aletras}.} \bibinfo{year}{2021}\natexlab{}.
\newblock \showarticletitle{Active learning by acquiring contrastive examples}.
\newblock \bibinfo{journal}{\emph{arXiv preprint arXiv:2109.03764}} (\bibinfo{year}{2021}).
\newblock


\bibitem[Marton(2014)]%
        {marton2014necessary}
\bibfield{author}{\bibinfo{person}{Ference Marton}.} \bibinfo{year}{2014}\natexlab{}.
\newblock \bibinfo{booktitle}{\emph{Necessary conditions of learning}}.
\newblock \bibinfo{publisher}{Routledge}.
\newblock


\bibitem[McHugh(2012)]%
        {mchugh2012interrater}
\bibfield{author}{\bibinfo{person}{Mary~L McHugh}.} \bibinfo{year}{2012}\natexlab{}.
\newblock \showarticletitle{Interrater reliability: the kappa statistic}.
\newblock \bibinfo{journal}{\emph{Biochemia medica}} \bibinfo{volume}{22}, \bibinfo{number}{3} (\bibinfo{year}{2012}), \bibinfo{pages}{276--282}.
\newblock


\bibitem[Monarch(2021)]%
        {monarch2021human}
\bibfield{author}{\bibinfo{person}{Robert~Munro Monarch}.} \bibinfo{year}{2021}\natexlab{}.
\newblock \bibinfo{booktitle}{\emph{Human-in-the-Loop Machine Learning: Active learning and annotation for human-centered AI}}.
\newblock \bibinfo{publisher}{Simon and Schuster}.
\newblock


\bibitem[Neuh{\"a}user(2011)]%
        {neuhauser2011wilcoxon}
\bibfield{author}{\bibinfo{person}{Markus Neuh{\"a}user}.} \bibinfo{year}{2011}\natexlab{}.
\newblock \bibinfo{title}{Wilcoxon-Mann-Whitney Test.}
\newblock
\newblock


\bibitem[No{\"e}(2022)]%
        {noe2022learning}
\bibfield{author}{\bibinfo{person}{Alva No{\"e}}.} \bibinfo{year}{2022}\natexlab{}.
\newblock \bibinfo{booktitle}{\emph{Learning to Look: Dispatches from the Art World}}.
\newblock \bibinfo{publisher}{Oxford University Press}.
\newblock


\bibitem[Quteineh et~al\mbox{.}(2020)]%
        {quteineh2020textual}
\bibfield{author}{\bibinfo{person}{Husam Quteineh}, \bibinfo{person}{Spyridon Samothrakis}, {and} \bibinfo{person}{Richard Sutcliffe}.} \bibinfo{year}{2020}\natexlab{}.
\newblock \showarticletitle{Textual data augmentation for efficient active learning on tiny datasets}. In \bibinfo{booktitle}{\emph{Proceedings of the 2020 Conference on Empirical Methods in Natural Language Processing (EMNLP)}}. \bibinfo{pages}{7400--7410}.
\newblock


\bibitem[Ramos et~al\mbox{.}(2020)]%
        {ramos2020interactive}
\bibfield{author}{\bibinfo{person}{Gonzalo Ramos}, \bibinfo{person}{Christopher Meek}, \bibinfo{person}{Patrice Simard}, \bibinfo{person}{Jina Suh}, {and} \bibinfo{person}{Soroush Ghorashi}.} \bibinfo{year}{2020}\natexlab{}.
\newblock \showarticletitle{Interactive machine teaching: a human-centered approach to building machine-learned models}.
\newblock \bibinfo{journal}{\emph{Human--Computer Interaction}} \bibinfo{volume}{35}, \bibinfo{number}{5-6} (\bibinfo{year}{2020}), \bibinfo{pages}{413--451}.
\newblock


\bibitem[Robeer et~al\mbox{.}(2021)]%
        {robeer2021generating}
\bibfield{author}{\bibinfo{person}{Marcel Robeer}, \bibinfo{person}{Floris Bex}, {and} \bibinfo{person}{Ad Feelders}.} \bibinfo{year}{2021}\natexlab{}.
\newblock \showarticletitle{Generating realistic natural language counterfactuals}. In \bibinfo{booktitle}{\emph{Findings of the Association for Computational Linguistics: EMNLP 2021}}. \bibinfo{pages}{3611--3625}.
\newblock


\bibitem[Schumann and Rehbein(2019)]%
        {schumann2019active}
\bibfield{author}{\bibinfo{person}{Raphael Schumann} {and} \bibinfo{person}{Ines Rehbein}.} \bibinfo{year}{2019}\natexlab{}.
\newblock \showarticletitle{Active learning via membership query synthesis for semi-supervised sentence classification}. In \bibinfo{booktitle}{\emph{Proceedings of the 23rd conference on computational natural language learning (CoNLL)}}. \bibinfo{pages}{472--481}.
\newblock


\bibitem[Settles(2009)]%
        {settles2009active}
\bibfield{author}{\bibinfo{person}{Burr Settles}.} \bibinfo{year}{2009}\natexlab{}.
\newblock \showarticletitle{Active learning literature survey}.
\newblock  (\bibinfo{year}{2009}).
\newblock


\bibitem[shen(2024)]%
        {shen2024Towards}
\bibfield{author}{\bibinfo{person}{Knearem shen}.} \bibinfo{year}{2024}\natexlab{}.
\newblock \showarticletitle{Towards Bidirectional Human-AI Alignment: A Systematic Review for Clarifications, Framework, and Future Directions}.
\newblock \bibinfo{journal}{\emph{URL http://arxiv.org/abs/2406.09264}} (\bibinfo{year}{2024}).
\newblock


\bibitem[Shrestha et~al\mbox{.}(2023)]%
        {shrestha2023exploratory}
\bibfield{author}{\bibinfo{person}{Rajesh Shrestha}, \bibinfo{person}{Omeed Habibelahian}, \bibinfo{person}{Arash Termehchy}, {and} \bibinfo{person}{Paolo Papotti}.} \bibinfo{year}{2023}\natexlab{}.
\newblock \showarticletitle{Exploratory Training: When Annonators Learn About Data}.
\newblock \bibinfo{journal}{\emph{Proceedings of the ACM on Management of Data}} \bibinfo{volume}{1}, \bibinfo{number}{2} (\bibinfo{year}{2023}), \bibinfo{pages}{1--25}.
\newblock


\bibitem[Suh et~al\mbox{.}(2019)]%
        {suh2019anchorviz}
\bibfield{author}{\bibinfo{person}{Jina Suh}, \bibinfo{person}{Soroush Ghorashi}, \bibinfo{person}{Gonzalo Ramos}, \bibinfo{person}{Nan-Chen Chen}, \bibinfo{person}{Steven Drucker}, \bibinfo{person}{Johan Verwey}, {and} \bibinfo{person}{Patrice Simard}.} \bibinfo{year}{2019}\natexlab{}.
\newblock \showarticletitle{Anchorviz: Facilitating semantic data exploration and concept discovery for interactive machine learning}.
\newblock \bibinfo{journal}{\emph{ACM Transactions on Interactive Intelligent Systems (TiiS)}} \bibinfo{volume}{10}, \bibinfo{number}{1} (\bibinfo{year}{2019}), \bibinfo{pages}{1--38}.
\newblock


\bibitem[Swati~Mishra(2023)]%
        {Swati2023Human}
\bibfield{author}{\bibinfo{person}{Jeffrey M~Rzeszotarski Swati~Mishra}.} \bibinfo{year}{2023}\natexlab{}.
\newblock \showarticletitle{Human Expectations and Perceptions of Learning in Machine Teaching}.
\newblock  (\bibinfo{year}{2023}), \bibinfo{pages}{13--24}.
\newblock


\bibitem[Szymanski et~al\mbox{.}(2024a)]%
        {szymanski2024comparing}
\bibfield{author}{\bibinfo{person}{Annalisa Szymanski}, \bibinfo{person}{Simret~Araya Gebreegziabher}, \bibinfo{person}{Oghenemaro Anuyah}, \bibinfo{person}{Ronald~A Metoyer}, {and} \bibinfo{person}{Toby Jia-Jun Li}.} \bibinfo{year}{2024}\natexlab{a}.
\newblock \showarticletitle{Comparing Criteria Development Across Domain Experts, Lay Users, and Models in Large Language Model Evaluation}.
\newblock \bibinfo{journal}{\emph{arXiv preprint arXiv:2410.02054}} (\bibinfo{year}{2024}).
\newblock


\bibitem[Szymanski et~al\mbox{.}(2024b)]%
        {szymanski2024limitations}
\bibfield{author}{\bibinfo{person}{Annalisa Szymanski}, \bibinfo{person}{Noah Ziems}, \bibinfo{person}{Heather~A Eicher-Miller}, \bibinfo{person}{Toby Jia-Jun Li}, \bibinfo{person}{Meng Jiang}, {and} \bibinfo{person}{Ronald~A Metoyer}.} \bibinfo{year}{2024}\natexlab{b}.
\newblock \showarticletitle{Limitations of the LLM-as-a-Judge Approach for Evaluating LLM Outputs in Expert Knowledge Tasks}.
\newblock \bibinfo{journal}{\emph{arXiv preprint arXiv:2410.20266}} (\bibinfo{year}{2024}).
\newblock


\bibitem[Taneja et~al\mbox{.}(2022)]%
        {taneja2022human}
\bibfield{author}{\bibinfo{person}{Karan Taneja}, \bibinfo{person}{Harshvardhan Sikka}, {and} \bibinfo{person}{Ashok Goel}.} \bibinfo{year}{2022}\natexlab{}.
\newblock \showarticletitle{Human-AI Interaction Design in Machine Teaching}.
\newblock \bibinfo{journal}{\emph{arXiv preprint arXiv:2206.05182}} (\bibinfo{year}{2022}).
\newblock


\bibitem[Williams and Moser(2019)]%
        {williams2019art}
\bibfield{author}{\bibinfo{person}{Michael Williams} {and} \bibinfo{person}{Tami Moser}.} \bibinfo{year}{2019}\natexlab{}.
\newblock \showarticletitle{The art of coding and thematic exploration in qualitative research}.
\newblock \bibinfo{journal}{\emph{International management review}} \bibinfo{volume}{15}, \bibinfo{number}{1} (\bibinfo{year}{2019}), \bibinfo{pages}{45--55}.
\newblock


\bibitem[Wisniewski and Bassok(1999)]%
        {wisniewski1999makes}
\bibfield{author}{\bibinfo{person}{Edward~J Wisniewski} {and} \bibinfo{person}{Miriam Bassok}.} \bibinfo{year}{1999}\natexlab{}.
\newblock \showarticletitle{What makes a man similar to a tie? Stimulus compatibility with comparison and integration}.
\newblock \bibinfo{journal}{\emph{Cognitive psychology}} \bibinfo{volume}{39}, \bibinfo{number}{3-4} (\bibinfo{year}{1999}), \bibinfo{pages}{208--238}.
\newblock


\bibitem[Wu et~al\mbox{.}(2021)]%
        {wu2021polyjuice}
\bibfield{author}{\bibinfo{person}{Tongshuang Wu}, \bibinfo{person}{Marco~Tulio Ribeiro}, \bibinfo{person}{Jeffrey Heer}, {and} \bibinfo{person}{Daniel~S Weld}.} \bibinfo{year}{2021}\natexlab{}.
\newblock \showarticletitle{Polyjuice: Generating counterfactuals for explaining, evaluating, and improving models}.
\newblock \bibinfo{journal}{\emph{arXiv preprint arXiv:2101.00288}} (\bibinfo{year}{2021}).
\newblock


\bibitem[Yan et~al\mbox{.}(2022)]%
        {yan2022concept}
\bibfield{author}{\bibinfo{person}{Litao Yan}, \bibinfo{person}{Miryung Kim}, \bibinfo{person}{Bjoern Hartmann}, \bibinfo{person}{Tianyi Zhang}, {and} \bibinfo{person}{Elena~L Glassman}.} \bibinfo{year}{2022}\natexlab{}.
\newblock \showarticletitle{Concept-annotated examples for library comparison}. In \bibinfo{booktitle}{\emph{Proceedings of the 35th Annual ACM Symposium on User Interface Software and Technology}}. \bibinfo{pages}{1--16}.
\newblock


\bibitem[Yang and Loog(2016)]%
        {yang2016active}
\bibfield{author}{\bibinfo{person}{Yazhou Yang} {and} \bibinfo{person}{Marco Loog}.} \bibinfo{year}{2016}\natexlab{}.
\newblock \showarticletitle{Active learning using uncertainty information}. In \bibinfo{booktitle}{\emph{2016 23rd International Conference on Pattern Recognition (ICPR)}}. IEEE, \bibinfo{pages}{2646--2651}.
\newblock


\bibitem[Zhang and Soergel(2016)]%
        {zhang2016process}
\bibfield{author}{\bibinfo{person}{Pengyi Zhang} {and} \bibinfo{person}{Dagobert Soergel}.} \bibinfo{year}{2016}\natexlab{}.
\newblock \showarticletitle{Process patterns and conceptual changes in knowledge representations during information seeking and sensemaking: A qualitative user study}.
\newblock \bibinfo{journal}{\emph{Journal of Information Science}} \bibinfo{volume}{42}, \bibinfo{number}{1} (\bibinfo{year}{2016}), \bibinfo{pages}{59--78}.
\newblock


\bibitem[Zhang et~al\mbox{.}(2017)]%
        {zhang2017research}
\bibfield{author}{\bibinfo{person}{Shengnan Zhang}, \bibinfo{person}{Yan Hu}, {and} \bibinfo{person}{Guangrong Bian}.} \bibinfo{year}{2017}\natexlab{}.
\newblock \showarticletitle{Research on string similarity algorithm based on Levenshtein Distance}. In \bibinfo{booktitle}{\emph{2017 IEEE 2nd Advanced Information Technology, Electronic and Automation Control Conference (IAEAC)}}. IEEE, \bibinfo{pages}{2247--2251}.
\newblock


\bibitem[Zhang et~al\mbox{.}(2023)]%
        {zhang2023peanut}
\bibfield{author}{\bibinfo{person}{Zheng Zhang}, \bibinfo{person}{Zheng Ning}, \bibinfo{person}{Chenliang Xu}, \bibinfo{person}{Yapeng Tian}, {and} \bibinfo{person}{Toby Jia-Jun Li}.} \bibinfo{year}{2023}\natexlab{}.
\newblock \showarticletitle{PEANUT: A Human-AI Collaborative Tool for Annotating Audio-Visual Data}. In \bibinfo{booktitle}{\emph{Proceedings of the 36th Annual ACM Symposium on User Interface Software and Technology}}. \bibinfo{pages}{1--18}.
\newblock


\bibitem[Zhang et~al\mbox{.}(2022)]%
        {zhang2022survey}
\bibfield{author}{\bibinfo{person}{Zhisong Zhang}, \bibinfo{person}{Emma Strubell}, {and} \bibinfo{person}{Eduard Hovy}.} \bibinfo{year}{2022}\natexlab{}.
\newblock \showarticletitle{A survey of active learning for natural language processing}.
\newblock \bibinfo{journal}{\emph{arXiv preprint arXiv:2210.10109}} (\bibinfo{year}{2022}).
\newblock


\end{thebibliography}
